\PassOptionsToPackage{pdftex}{graphicx}
\PassOptionsToPackage{pdftex}{hyperref}

\documentclass[submit]{epsv8}

\usepackage{graphicx}
\usepackage{hyperref}
\usepackage{amsmath}

\title{Growth and Thermal Evolution of Icy Planetesimals}
\author{
Jun Kimura, Department of Earth and Space Sciences,
The University of Osaka, Osaka 560-0043, Japan,
junkim@ess.sci.osaka-u.ac.jp\\
Ryusei Satoh, The University of Osaka, rsatoh@ess.sci.osaka-u.ac.jp\\
Kentaro Terada, The University of Osaka, teraken@ess.sci.osaka-u.ac.jp\\
Sho Sasaki, The University of Osaka, sasakisho@ess.sci.osaka-u.ac.jp
}

\abstract{
Icy planetesimals are thought to contribute to the volatile inventory of terrestrial planets and serve as building blocks of icy bodies in the outer Solar System.
Samples from the C-type asteroid Ryugu, collected by the Hayabusa-2 spacecraft, indicate a low-temperature history with aqueous alteration and organic materials.
In contrast, iron meteorites with isotopic ratios similar to those of carbonaceous chondrites suggest exposure to higher temperatures.
These findings imply that the thermal evolution of icy planetesimals is highly diverse. 
Since direct exploration provides only localized data, understanding this diversity requires comparing observational results with model calculations that incorporate key evolutionary processes.
We develop a model including radial growth, impact heating, water phase changes, aqueous alteration, and structural differentiation, to re-evaluate the thermal evolution of icy planetesimals during the first 100 Myr after the formation of Calcium--Aluminum--rich Inclusions (CAIs).
The model considers final radius (10--1000 km), timing of growth onset (1.0 or 2.0 Myr after CAI), growth duration (0.4 or 4.0 Myr), and growth mode (linear or runaway).
Our results show that larger planetesimals generally reach higher temperatures, but growth timing and mode significantly affect thermal evolution. Early accretion leads to higher temperatures, with some bodies reaching the Fe--FeS eutectic (1250 K), while delayed or prolonged growth reduces heating.
Our results show that the constituent materials of Ryugu, which kept below 40$^{\circ}$C, likely formed near the surface of a hydrated mineral layer. 
This is possible even in planetesimals several hundred kilometers in size due to efficient heat transport via convection.
If accretion begins 2.0 Myr after CAI and completes in 0.4 Myr, a wide region in such a body could yield Ryugu's material.
Evolution into a 250 km body with a 170--200 km hydrous core and overlying liquid water layer may resemble Saturnian icy moon Enceladus.
For a later onset and longer duration of growth, even aqueous alteration could be prevented.
In contrast, metal melting in the deep regions of rapidly formed icy planetesimals larger than 200\,km could originate iron meteorites with isotopic signatures similar to those of carbonaceous chondrites.
}


\begin{document}

\maketitle

\section{Introduction}
Understanding the evolution of icy planetesimals is crucial for elucidating the overall history of the Solar System and advancing astrobiology.
These bodies are considered not only the fundamental building blocks of icy bodies but also potential sources of water and prebiotic materials delivered to Earth \citep[e.g.][]{sill+1978, owen+1995, raymond+2004}.
However, direct exploration of icy planetesimals via close-range observations or sample return missions has yet to be conducted. 
As a result, their thermal evolution has been inferred through a combination of theoretical models and analyses of asteroidal materials and meteorites, which are considered remnants of icy planetesimals. 

Meteorites are generally classified into two groups, non-carbonaceous chondrites (NCs) and carbonaceous chondrites (CCs), based on their isotopic compositions. 
CCs are thought to have formed beyond the snow line where icy materials can condense. 
Recent isotopic studies suggest that some iron meteorites fall into the CC group rather than the NC group \citep[e.g.][]{kruijer2017, kruijer2020}.
Carbonaceous chondrites (CCs) contain organic matter, volatile components, and minerals indicative of aqueous alteration. 
During aqueous alteration, anhydrous minerals transform into hydrous phyllosilicates (e.g., olivine into serpentine). 
Metallic iron oxidizes into magnetite and other iron oxides or hydroxides, and carbonate minerals precipitate during this process.
The presence of hydrous phyllosilicates, magnetite, and carbonates in CCs suggests that their parent bodies were icy planetesimals rich in water and formed at low temperatures.
Small bodies exhibiting spectral characteristics similar to those of CCs in the visible and infrared regions are classified as C-type asteroids, which are believed to have originated from icy planetesimals that underwent similar thermal evolution \citep[e.g.][]{bottke2006, demeo2014}. 

The C-type asteroid Ryugu (162173) has been extensively studied due to its spectral characteristics.
Its spectrum exhibits absorption bands at 2.7\,$\mu$m, indicative of hydrous minerals, and at 3.4\,$\mu$m, associated with organic matter and carbonate minerals, resembling the spectral features of CCs \citep{bates+2020, yada+2022, pilorget+2022}. 
Analysis of samples retrieved from Ryugu by the Hayabusa2 spacecraft revealed strong similarities in elemental and mineral compositions \citep{yokoyama+2023, nakamuraT+2023, ito+2022, nakamuraE+2022, okazaki+2022}, as well as in isotopic compositions of O, Ca, Ti, Cr, and Fe \citep{yokoyama+2023, nakamuraE+2022, hopp+2022, greenwood+2023} with CCs, particularly CI chondrites, which closely resemble the solar composition. 
Furthermore, Ryugu samples contain serpentine, magnetite, carbonates, and organic matter indicative of aqueous alteration \citep{ito+2022, nakamuraE+2022, yokoyama+2023, nakamuraT+2023}, supporting the hypothesis that the parent bodies of Ryugu and CCs were icy planetesimals.

The water-to-rock mass ratio (W/R) during aqueous alteration has been estimated to range from 0.2 to 0.9 \citep{nakamuraT+2023}. 
Chronological analyses using Mn-Cr dating suggest that aqueous alteration occurred within a few Myr after the formation of CAIs, although different studies report somewhat varying ages \citep{nakamuraE+2022,yokoyama+2023,mccain2023,sugawara+2024}.
Temperature estimates derived from oxygen isotope analysis of carbonate minerals suggest 37\,$\pm$\,10\,${}^\circ$C \citep{yokoyama+2023}, while pentlandite lattice spacing and chemical composition indicate 20\,$\pm$\,29.5\,${}^\circ$C \citep{nakamuraT+2023}. 
Additionally, studies of aliphatic compounds in the Murchison meteorite, although belonging to the CM group, provide an additional constraint on the thermal conditions of carbonaceous chondrites in general, suggesting a temperature below 30\,${}^\circ$C \citep{ito+2022}.
These findings impose constraints on the thermal evolution of the parent body.

It is possible to infer Ryugu's parent body from these estimates of orbital observations, numerical calculations, calculations, the size, structure, and formation time.
\citet{walsh+2013} suggested that Ryugu may belong to either the Eulalia or Polana asteroid family based on orbital observations, but they argued that the Eulalia family provides a better match. 
They estimated the radius of the Eulalia parent body to be approximately 50~km, considering the current total mass of Eulalia family objects and the estimated mass lost over time.
\citet{mccain2023} estimated the radius of the parent body to be $\sim$20~km and constrained its accumulation age to within 1.8~Myr after the formation of CAIs, noting this as an upper bound. Using the parent body radius estimated by \citet{walsh+2013}, \citet{kitazato+2019}, and aqueous alteration estimates by \citet{yokoyama+2023}, \citet{nakamuraT+2023} calculated the formation time of the undifferentiated Ryugu parent body to be between 1.8~Myr (for W/R\,=\,0.2) and 2.9~Myr (for W/R\,=\,0.9) after CAIs, thereby providing a plausible range that reflects model assumptions.

Other modeling approaches have also addressed Ryugu's parent body: \citet{neumann2021} suggested that only very small parent bodies, with radii smaller than $\sim$10\,km, could retain high porosity, requiring early accretion within $\sim$2--3 Myr after CAIs; 
\citet{tang2023} considered bodies smaller than $\sim$50\,km and found that Ryugu--like alteration could only be reproduced for early accretion times within $\sim$1--3 Myr after CAIs. 
Taken together with the estimates by \citet{walsh+2013}, \citet{mccain2023}, \citet{kitazato+2019}, and Nakamura et al. (2023), these results indicate that a wide range of scenarios, both in size and accretion time, have been proposed for Ryugu's origin. 
In addition, numerous studies have modeled hydrated or water--rich planetesimals more generally, including Ceres-like bodies \citep{neumann2015, neumann2020}, parent bodies of carbonaceous chondrites with possible core formation \citep{ma2022,neumann2024}, general porosity evolution \citep{henke2012a, henke2012b, henke2013, neumann2014}, and serpentinization-driven processes in icy satellites and Kuiper Belt objects \citep{malamud2013, malamud2015}.

In addition, \citet{neveu2015} developed the IcyDwarf code, which simultaneously treats thermal evolution and complex geochemical reactions such as serpentinization, dehydration, and porosity evolution. 
The code has been applied to various icy bodies, including Ceres and Saturn's mid-sized moons, providing quantitative constraints on water-rock ratios, hydration state, and volume changes during alteration. 
Similar modeling was extended to the large Kuiper Belt object Haumea by \citet{noviello2022}, who found that its rocky core experienced sequential hydration and dehydration reactions under comparable temperature ranges to those studied here. 
These works represent important geochemical complements to our present model, which focuses on the coupled effects of accretional growth, impact heating, and conductive-convective heat transport.

While the existence of low-temperature icy planetesimals, such as Ryugu's parent body, has been suggested, iron meteorites with isotopic compositions similar to those of carbonaceous chondrites (CCs) have been discovered \citep[e.g.][]{kruijer2017, nanne+2019}.
This implies that some planetesimals have experienced temperatures high enough for iron to melt and undergo differentiation.
Primary isotopic studies indicate that iron meteorites and achondrites accreted approximately 1.0~Myr and 2.0~Myr after CAI formation, respectively, while chondrites accreted around 2.5~Myr after CAIs \citep{kruijer2014,budde2018,kruijer2017}.
Although it cannot be said with certainty that iron meteorites with isotopic compositions similar to those of CCs are of icy planetesimal origin \citep{kruijer2017,kruijer2020}, these findings suggest that icy planetesimals exhibit a thermally diverse evolutionary history, ranging from low temperatures of several tens of degrees Celsius to high temperatures sufficient for metallic differentiation.
However, a comprehensive model describing the evolution of such icy planetesimals has yet to be established. 
Furthermore, information derived from meteorite and sample analyses represents only a fragment of the parent body, making it challenging to generalize findings to the entire parent body.
To fully understand the internal structure and evolution of icy planetesimals and other bodies, it is essential to develop an evolutionary model that integrates the various fundamental processes occurring within them. By doing so, the results of sample analyses can be contextualized within a broader framework of parent body evolution.

The internal thermal evolution of primordial bodies during solar system formation has been extensively investigated through theoretical studies. 
The primary heat source during this period was the decay of short-lived radionuclides, particularly $^{26}$Al (half-life: 0.72\,Myr), which was proposed by \citet{urey1955} as the dominant heat source. 
\citet{lee+1977} were the first to report an initial $^{26}$Al/$^{27}$Al ratio of approximately 5$\times$10$^{-5}$, based on elemental measurements of the Allende meteorite, a CV chondrite. 
\citet{minster+1979} used rubidium-strontium ($^{87}$Rb-$^{87}$Sr) dating of the Guare\~{n}a meteorite, an H chondrite, to demonstrate that H chondrites originated from the primordial solar nebula. 
They estimated the age of agglomeration of the H-chondrite parent body's core (central region) to be 4518\,$\pm$\,26~Myr and, using constraints derived from a model incorporating maximum temperature and cooling rate, and $^{26}$Al as the primary heat source, proposed that the H chondrite parent body had a size of 150-175~km.
\citet{miyamoto+1980} assumed the surface temperatures of H and L chondrite parent bodies, which belong to the category of ordinary chondrites, to be 1300\,K. 
Using a simple calculation based on the Fourier number ($k\Delta t/R^2$), an index of heat transport efficiency by thermal conduction, they estimated the radius of the ordinary chondrite parent body to be 2-10\,km.
Subsequently, \citet{miyamoto+1981} conducted the first numerical calculations of the internal thermal evolution of planetesimals. 
By employing an infinite series to compute temperature variations due to internal heat generation and thermal conduction, they proposed that the initial temperatures of the H and L chondrite parent bodies were 200\,K and 180\,K, respectively, and that both had radii of 85\,km.
Since then, advancements in sample analysis techniques and computational power have led to numerous numerical simulations of planetesimals and parent bodies, significantly improving our understanding of their thermal evolution.

However, most of these studies have focused on the thermal evolution of rocky planetesimals, while relatively few have examined the thermal evolution of icy planetesimals.
This is primarily due to the complexity of fundamental processes involved, such as uncertainties in the physical properties arising from the coexistence of rock and water, aqueous alteration, dehydration reactions, heat transport by thermal convection in addition to conduction, accretionary growth, and the simultaneous differentiation of components leading to the formation of layered structures. In previous numerical models, these processes have only been incorporated individually.
Several studies have addressed specific aspects of these processes. \citet{grimm+1989, young+1999, wakita+2011, bland+2017} investigated the melting of ice and aqueous alteration due to temperature increases inside carbonaceous chondrite parent bodies. 
\citet{lichtenberg+2016} and \citet{merk+2002} examined thermal convection in two- and three-dimensional models of the internal thermal evolution of rocky planetesimals, while \citet{merk+2002} also considered the growth of planetesimals through the accretion of rocky planetesimals. 
\citet{sramek+2012} and \citet{jutzi+2020} analyzed impact heating as a heat source for rocky and icy planetesimals, respectively. 
\citet{wakita+2011} investigated the internal differentiation of water and rock due to heating, whereas \citet{bland+2017} examined changes in porosity and particle transport.
However, we emphasize that these studies considered each process in isolation. A comprehensive investigation that integrates multiple interacting processes to explore the overall thermal evolution of icy planetesimals remains lacking.
In addition, several modeling studies have examined hydrated or water-rich bodies such as Ceres, carbonaceous chondrite parent bodies, and Ryugu itself.
These works have offered important insights into accretion, porosity evolution, and internal differentiation, but they were designed as case studies of particular bodies or processes. What is still missing is a systematic parametric exploration of undifferentiated icy planetesimals across a range of water-to-rock ratios, particularly at scales relevant to Ryugu-like bodies.
The objective of this study is to address this specific gap.

The objective of this study is to develop an evolutionary model that accounts for internal heating, thermal transport via conduction and convection, ice melting, aqueous alteration, dehydration processes, and differentiation-driven structural changes within icy planetesimals. 
Additionally, the model incorporates accretionary growth and impact heating, providing a re-evaluation of their internal thermal evolution. 
Numerical simulations are conducted to explore the thermal evolution of icy planetesimals up to 100~Myr, focusing on the early phase dominated by the decay of short-lived radionuclides such as $^{26}$Al. 
The models consider variations in final radius, onset timing of planetesimal growth, duration of accretional growth, and growth mode, under the assumption of an initial temperature of 70~K and an initial radius of 1~km. 
We note that possible longer-term processes beyond 100~Myr, such as the freezing of liquid water layers and associated geophysical changes, are outside the scope of this study.
This paper presents the model framework and assumptions (Chapter 2), summarizes parameter study results (Chapter 3), discusses implications for the diversity of icy planetesimal thermal evolution and parent bodies (Chapter 4), and concludes with final remarks (Chapter 5).
\clearpage
\section{Methods}
\subsection{Model overview for the planetesimal evolution}\label{method_overview}
\begin{figure}[p]
\begin{center}   
\includegraphics[width=1.0\textwidth]{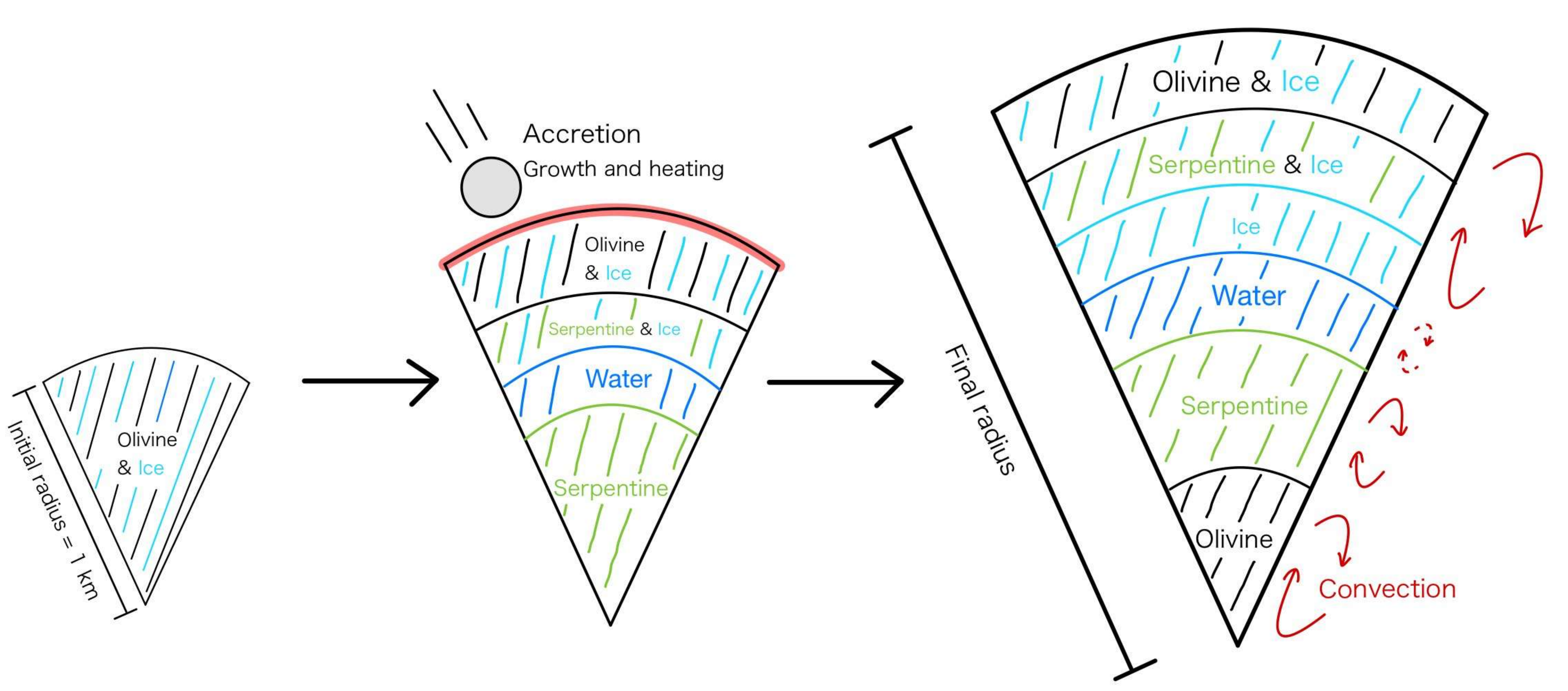} 
\end{center} 
\caption{Schematic diagram of the interior evolution model in this work. 
Initial planetesimal with radius of 1\,km grows by accretion.
Eventually, six layered structures in maximum are formed according to its thermal state.
Heat transfer rate due to the solid-state convection is calculated separately for the anhydrous mineral (olivine) layer, the hydrated mineral (serpentine) layer, and the ice and initial composition layer.
The water mantle is assumed to be isothermal.}
\label{inner_structure}
\end{figure}

To describe the process by which an initial 1 km-radius icy planetesimal undergoes structural and thermal evolution during its growth through collisional accretion, we have developed a numerical model.
In this model, up to five stages of structural transformation occur depending on the internal thermal state.
The extent to which these stages progress depends on the thermal evolution.
In this study, the term "core" denotes the central region of the body occupied by rocky material, irrespective of its hydration state or the presence of metallic melt.
\begin{enumerate}
\setcounter{enumi}{-1} 
\item Initial Composition and Accretionary Material: The radius of the body changes over time due to accretion. The initial composition of the body and the accreting material (hereafter referred to as accretionary material) consist of a mixture of ice and anhydrous minerals (olivine).

\item Aqueous Alteration: As the temperature rises and ice melts, aqueous alteration occurs instantaneously, transforming olivine into hydrous minerals and leading to the formation of a mixture of hydrous minerals and excess liquid water (see Section 2.6).

\item Differentiation of Hydrous Minerals and Water Mantle: Following stage (1), the presence of hydrous minerals and excess liquid water from aqueous alteration results in density-driven segregation, where hydrous minerals sink to the center, forming a core, while liquid water accumulates above the core, creating a liquid water layer (water mantle).

\item Dehydration Reaction: A further increase in temperature triggers dehydration reactions, producing anhydrous minerals (olivine) and liquid water. 
The anhydrous minerals migrate toward the center due to density differences, further densifying and restructuring the existing central rocky region (hereafter referred to as the ''core''), while the liquid water generated by dehydration ascends and is incorporated into the existing water mantle.

\item Solidification of the Water Mantle: As the interior of the body cools, the water mantle solidifies, forming an ice layer.
\end{enumerate}
Here we modelled the anhydrous minerals as olivine which is a mineral commonly used to characterize a planetary rocks \citep{sohl02}, and the hydrous minerals as serpentine which is considered a prevalent hydrous mineral in the interiors of planetary bodies \citep{vance18}.
We note that the onset temperature for dehydration and the threshold for metallic melting are specified in Section 2.6. 
In regions where metallic melting occurs, metal-silicate differentiation subsequently follows.
In the case of the highest-temperature evolution, the body differentiates into six layers from the center outward: 
Layer\#1(L1): A central core composed of olivine, L2: An outer core composed of serpentine, L3: A water mantle, L4: An ice layer, L5: A layer of liquid water, ice, and serpentine, and L6: An outermost layer with the initial composition of ice and olivine (Figure\,\ref{inner_structure}).

\subsection{Initial state of planetesimals}
In this study, the W/R for the initial composition of the planetesimal and for the accreting materials are both set to 0.5, based on the W/R\,=\,0.2--0.9 obtained from analyses of Ryugu samples \citep{nakamuraT+2023}. 
We assume that water initially existed as solid ice, while all rock is in the form of olivine (Mg$_{0.5}$,Fe$_{0.5}$)$_{2}$SiO$_{4}$).
The initial temperature of the planetesimal is homogeneously set to 70\,K from the surface to the center, with the surface temperature fixed at 70\,K. 
Such assumption is based on the presence of CO$_{2}$ ice within pyrrhotite crystals in Ryugu samples, suggesting that the Ryugu parent body formed beyond the CO$_{2}$ snowline \citep{nakamuraT+2023}.
The adsorption temperature of CO$_{2}$ is 70 K \citep{suresh+2024}. 
According to the minimum mass disk model, temperature of the nebula, $T_{nebula}$, is described as
\begin{math}
T_{nebula}=280\left(R_{AU}/1\,AU\right)^{-0.5}, 
\end{math}
where $R_{AU}$ [AU] denotes the object's distance from the Sun.
This gives $\sim$70~K at $\sim$16~AU from the Sun \citep{hayashi1981}. 
We note that this temperature is adopted as a representative low-temperature condition for icy planetesimal modeling, and does not imply that Ryugu's parent body or the Eulalia family members formed at this location.
We adopt an initial radius of 1~km, which subsequently grows to various final sizes through accretion. 
This initial size was selected to reduce sensitivity to thermal effects during the earliest stages of evolution.
Using $t\sim R^{2}/\kappa$ with thermal diffusivity $\kappa\sim10^{-6}$\,m$^{2}$\,s$^{-1}$, the thermal diffusion time, $t$, for 1 km is about $3\times10^{4}$ years, shorter than the $^{26}$Al half-life and far shorter than the multi-Myr accretion times considered. 
Therefore this choice confines the initial effect to a short transient and is not expected to materially affect the conclusions.
%
\subsection{Numerical model for thermal evolution}

The following is the general equation for heat transfer:
\begin{equation}
\rho C_{p}\frac{dT}{dt}=-\frac{1}{r^{2}}\frac{d}{dr}\left(r^{2}F_{cond}+r^{2}F_{conv}\right)+\rho Q
\label{eq:heat_transfer}
\end{equation}
where $\rho$ is density, $C_{p}$ is specific heat, $T$ is temperature, $t$ is time, $r$ is radial distance from the centre, and $Q$ is heat production rate per unit mass.
The conductive and convective heat fluxes are given by
\begin{equation}
F_{cond}=-k_{c}\frac{dT}{dr},
\end{equation}
\begin{equation}
F_{conv}=-k_{v}\left\{\frac{dT}{dr}-\left(\frac{dT}{dr}\right)_{ad}\right\},
\end{equation}
where $k_{c}$ is thermal conductivity, and $k_{v}$ is {\it effective} thermal conductivity, which emulates the effect of thermal convection.
$k_{v}$ is expressed as follows \citep[e.g.,][]{abe93,kimura+2009}:
\begin{equation}
\!k_{v}\!=\!\left\{
\begin{array}{ll}
\displaystyle{-\frac{\rho^{2}C_{p}\alpha g\ell^{4}}{18\eta}
\left\{\frac{\partial T}{\partial r}
\!-\!\left(\frac{\partial T}{\partial r}\right)_{ad}\right\}}
& \textrm{for }\;\displaystyle{\frac{\partial T}{\partial r}
\!<\!\left(\frac{\partial T}{\partial r}\right)_{ad}} \\
0 & \textrm{for }\;\displaystyle{\frac{\partial T}{\partial r}
\!>\!\left(\frac{\partial T}{\partial r}\right)_{ad}}
\end{array}
\right.
\end{equation}
where $g$ is the gravitational acceleration, $\alpha$ is thermal expansion coefficient, and $\eta$ is viscosity.
$\ell$ denotes a characteristic length scale (mixing length), introduced to evaluate convective thermal transport. 
A more detailed explanation is provided later.
($dT/dr)_{ad}$ is the adiabatic temperature gradient given by
\begin{equation}
\left(\frac{dT}{dr}\right)_{ad}=-\frac{\alpha gT}{C_{p}}.
\end{equation}
Note that the both $dT/dr$ and $(dT/dr)_{ad}$ are negative; a positive convective heat flux appears only if the temperature gradient is steeper than the adiabatic temperature gradient.
$\ell$ linearly increases with depth to the peak value $b$ until it reaches the peak depth $a$, and then it decreases linearly with depth as follows:
\begin{equation}
\!\ell\!=\!\left\{
\begin{array}{ll}
\displaystyle{\frac{b}{a}(R_{top}-r)}
& \textrm{for }\;\displaystyle{r\!\ge\! R_{top}-aD} \\
\displaystyle{\frac{b}{1-a}(r-R_{bot})}
& \textrm{for }\;\displaystyle{r\!\le\! R_{top}-aD}
\end{array}
\right.
\end{equation}
where $R_{bot}$ and $R_{top}$ are the radii at the bottom and top of the convective layer, respectively; 
$D$ is the thickness of the layer.
For the deeper olivine core and surrounding serpentine layer (Figure\,\ref{inner_structure}), $a=0.5$ and $b=0.5$, where $\ell$ is assumed to represent the distance to the nearest boundary of the layer to reproduce a $Nu\sim Ra^{1/3}$ relationship \citep[e.g.][]{sasaki86,abe93,senshu+2002,kimura+2009}.
For other outer solid layers having a large curvature, that is, $f=R_{bot}/R_{top}>0.5$, the following modifying values of $a$ and $b$ are well consistent with the predictions obtained by the 3D calculations \citep{kamata18,kimura24}:
\begin{equation}
a(f,\gamma)=a_{2}(\gamma)f^{2}+a_{1}(\gamma)f+a_{0}(\gamma),
\end{equation}
\begin{equation}
b(f,\gamma)=b_{2}(\gamma)f^{2}+b_{1}(\gamma)f+b_{0}(\gamma),
\end{equation}
\begin{equation}
a_{2}(\gamma)=-41.2\exp(-0.297\gamma)-0.456,
\end{equation}
\begin{equation}
a_{1}(\gamma)=58.6\exp(-0.292\gamma)+0.704,
\end{equation}
\begin{equation}
a_{0}(\gamma)=-21.0\exp(-0.290\gamma)+0.624,
\end{equation}
\begin{equation}
b_{2}(\gamma)=3.96\exp(-0.167\gamma),
\end{equation}
\begin{equation}
b_{1}(\gamma)=-6.93\exp(-0.178\gamma),
\end{equation}
\begin{equation}
b_{0}(\gamma)=2.90\exp(-0.127\gamma),
\end{equation}
\begin{equation}
\gamma=\frac{2c^{2}_{0}\Delta T}{2c_{0}T_{b}+c_{1}-\sqrt{c^{2}_{1}+4c_{0}c_{1}T_{b}}},
\end{equation}
where $\Delta T$ is the temperature difference across the layer, and $T_{b}$ is the temperature at the base of the layer.
$c_{0}=1.23/f^{1.5}$ and $c_{1}=E_{a}/R_{g}$ where $E_{a}$ is activation energy and $R_{g}$ is the gas constant.

The viscosity of ice, $\eta_{ice}$, strongly affects the efficiency of heat transfer in the ice shell. The ice viscosity's large temperature dependency is well-approximated by
\begin{equation}
\eta_{ice}=\eta_{ice,ref}\exp{\left[\frac{E_{ice}}{R_g}\left(\frac{1}{T}-\frac{1}{T_{m}}\right)\right]}.
\end{equation}
where $\eta_{ice,ref}$ is the viscosity of ice at the melting temperature, $T_{m}$, $R_{g}$ is the gas constant, and $E_{ice} = 60$\,kJ\,mol$^{-1}$ \citep{goldsby+2001}.
A typical value of $\eta_{ice,ref}$ is in a range between 10$^{13}$ and 10$^{15}$ Pa\,s \citep[e.g.][]{hussmann15}, which is comparable with that of terrestrial glaciers, although it can vary largely depending on many parameters, such as grain size.
Mixing an irregular olivine particles into ice increases the viscosity \citep{yasui+2008}, thus we set $\eta_{ice,ref}$ to be $1.0 \times 10^{15}$ for ice Ih and initial composition.

The physical properties of olivine and serpentine, except for the viscosity, are assumed to be uniform.
We consider the temperature--dependent olivine viscosity given by
\begin{equation}
\eta_{oliv}=\eta_{oliv,ref}\exp{\left[A\frac{T_{solidus}}{T}\right]}
\end{equation}
where $\eta_{oliv,ref}$ is reference viscosity of the olivine, $A$ is a constant and $T_{solidus}$ is the solidus temperature of the olivine.
We adopted $\eta_{oliv,ref}$\,=\,4.9\,$\times$\,10$^{8}$ Pa\,s, $A$\,=\,23.25 and $T_{solidus}\,=\,1.0\times10^{-7}\times P$, where $P$ is the pressure \citep{zhang+1994}.
The viscosity of serpentine is described as follows:
\begin{equation}
\eta_{serp}=\eta_{serp,ref}\exp{\left[\frac{E_{serp}+PV^*}{nR_gT}\right]}
\end{equation}
where $\eta_{serp,ref}$\,=\,$1.5\times10^{20}$\,[Pa\,s] is the pre-exponential factor of the flow law and does not correspond to a viscosity at any specific temperature, $E_{serp}=17.6$\,[kJ], $V^{*}\,=\,3.2\times10^{-6}$\,[m$^{3}$], $n=3.8$ \citep{hilairet+2007}.

The density, specific heat, and thermal conductivity are calculated as average values based on the mass fraction of rock and water, considering their composition and state, as expressed in the following formula \citep{wakita+2011}.
The expansion coefficient of the initial composition is determined to be 9.8$\times$10$^{-5}$, calculated from the volume fractions of rock and ice. 
The density, specific heat, and thermal conductivity of rock and ice used in these calculations are listed in Table\,\ref{table_den_speh_cond}.
For the initial composition, bulk density, $\overline{\rho}$, bulk specific heat. $\overline{C_{p}}$, and bulk thermal conductivity, $\overline{k_{c}}$, can be calculated from the following equations:%
\begin{equation}\label{den_speh_cond}
\overline{\rho}\,=\,1/\left\{f_{mr}/\rho_r + \left(1-f_{mr}\right)/\rho_i\right\} \\
\end{equation}
\begin{equation}
\overline{C_p}\,=\,f_{mr}Cp_r + \left(1-f_{mr}\right)C_{pi} \\
\end{equation}
\begin{equation}
\overline{k_c}\,=\,f_{vr}k_{cr} + \left(1-f_{vr}\right)k_{ci}
\end{equation}
where the subscript $r$ and $i$ denote rock and ice, respectively.
The terms $f_{mr}$ and $f_{vr}$ represent the mass fraction and volume fraction of rock relative to the total composition, respectively, and are related by the equation $f_{vr}$\,=\,$f_{mr}\bar{\rho}/\rho_{r}$.
The same approach applies to ice.
For simplicity, the densities of both water and ice are assumed to be 1,000\,kg/m$^{3}$ \citep{wakita+2011}.
For the initial composition, $\overline{\rho}$\,=\,1,868\,kg/m$^{3}$, $\overline{C_p}$\,=\,1,240\,J/kg\,K, $\overline{k_c}$\,=\,2.50\,W/m\,K are derived.

The thermal expansion coefficient of olivine remains nearly constant around 800\,K \citep{suzuki1975}, while that of serpentine exhibits temperature dependence at lower temperatures \citep{osako+2010}.
The linear expansion coefficient of ice is nearly constant around 273\,K \citep{butkovich1959}. 
Therefore, the volume expansion coefficient of ice, hydrous rock and dehydrated rock are assumed to be constant and independent of temperature (Table\,\ref{table_den_speh_cond}).
%
\subsection{Radial growth of planetesimals through accretion}
Icy planetesimals are thought to have formed through the accumulation and growth of dust and small, volatile-rich bodies in the outer Solar System. 
Various hypotheses exist regarding the onset, duration, and mode of accretional growth, which influence the growth rate of their radii.

The prevailing theory suggests that small bodies in the protosolar disk, once they reach a mass of approximately 10$^{9}$--10$^{15}$\,kg (corresponding to a size of 0.1\,--\,10 km), undergo runaway growth. During this phase, they rapidly attract surrounding dust through gravitational interactions. Subsequently, they transition to oligarchic growth, where a small number of sufficiently large bodies continue to accumulate material and grow \citep[cf.][]{stewart+1988, wetherill+1989, kokubo+1998, sramek+2012}.
In this study, accretion is described by the radius growth function $R(t)$ with explicit starting period and duration. 
As a first-order description, a rapid early increase in $R$ models gravitational collapse, whereas sustained growth models pebble accretion. 
The thermal evolution in our model depends primarily on the final radius and on the accretion duration, rather than on the specific mechanism.

The time evolution of a growing body's mass and radius can be described by the following equation:

Here, $\alpha$ and $\beta$ are related by $\beta=3\alpha+1$. 
Runaway growth is characterized by $\alpha=1/3$ and $\beta=2$, whereas oligarchic growth corresponds to $\alpha=-1/3$ and $\beta=0$.
A larger $\beta$ value indicates that the planetesimal's radius increases more rapidly in the later stages of accretion. 
Consequently, radiogenic heating by $^{26}$Al is small only when accretion starts late and proceeds slowly, so that the body remains small while most $^{26}$Al decays. 
By contrast, an early onset and short accretion duration allows substantial $^{26}$Al heating. 
In this study we do not include long-lived radionuclides in our model, although their effects can be important for large bodies due to slow cooling.
However, impact heating is expected to have a significant effect (further details are discussed in Section 2.5).

In this study, two values of $\beta=0$ and $2$, are used as parameters for calculations.
For instance, in the case of $\beta=2$, an duration of accretion of 4.0\,Myr and a final radius of 1,000\,km, the time required for the radius to grow from 1\,km to 10\,km accounts for more than 90\% of the total accretion period. 
This indicates that the planetesimal's radius undergoes a very rapid increase in the final stages of growth.

\citet{kokubo+2000} performed N-body simulations of runaway growth using 3,000 objects with masses determined by the initial mass distribution and oligarchic growth using 4,000 such objects. 
Their calculations indicate that accretion times can span a broad range depending on dynamical conditions. In this study, we therefore examine two representative accretion durations, 0.4~Myr and 4.0~Myr, focusing on the regime where $^{26}$Al can affect thermal evolution.

Based on this, the duration of accretion in this study is set to two values: 0.4~Myr and 4.0~Myr. 
Planetesimal is assumed to start growing 1.0 or 2.0~Myr after the formation of CAIs, with an initial size of 1\,km and a final size ranging from 10~km to 1,000~km.
The parameters related to accretional growth used in the calculations are listed in Table\,\ref{table_values_properties}
.
%
\subsection{Heat sources}
The heat source $Q$ of a planetesimal consists of two components: the decay heat of radioactive isotopes ($Q_{decay}$) and the impact heating during accretion ($Q_{imp}$). 
We considers two short-lived radioactive isotopes, $^{26}$Al and $^{60}$Fe, as heat sources.

The decay heat generated by radioisotopes can be expressed as follows \citep{wakita+2011}:
\begin{equation}
Q_{decay}=\sum_r\left(\frac{E_r\lambda_rf_r}{m_r}\right)\left[\left(\frac{^{26}\mathrm{Al}}{^{27}\mathrm{Al}}\right)_0 \mathrm{or} \left(\frac{^{60}\mathrm{Fe}}{^{57}\mathrm{Fe}}\right)_0\right]\rho_{oliv}f_{vr}\exp({-\lambda_r t})
\end{equation}
where $r$ denotes the radioisotope, $E_r$ is the energy released per decay event, $m_r$ is the mass of a single nuclide, and $t$ is the time elapsed since the formation of CAIs.  
The total abundance of the isotope in the planetesimal is represented by $f_r$, and in this study, the values from CI chondrites are used. 
The decay constant, $\lambda_r$, is defined as $\lambda_r$\,=\,$\ln 2/T_{r1/2}$ where $T_{r1/2}$ is the half-life of the isotope. 
($^{26}$Al/$^{27}$Al)$_0$ or ($^{60}$Fe/$^{57}$Fe)$_0$ represents the initial isotope abundance ratio at the time of CAI formation. 
The density of olivine is denoted as $\rho_{oliv}$, and $f_{vr}$ is the volume fraction of olivine in each layer.

The physical properties of the radioisotopes are summarized in Table\,\ref{properties of radionuclide}.

For a growing planetary body, an impactor of mass $dm_{imp}$ brings gravitational energy $dE_{g}=(GM/R)d_{imp}$ where $M=(4/3)\pi\bar{\rho}R^{3}$ is the mass of the target (planetesimal).
The impactors add to the target mass, therefore, $dm_{imp}/{dt}=dM/dt=4\pi\bar{\rho}R^2\dot{R}$, and the gravitational energy released per unit time is given as follows \citep{sramek+2012},
\begin{equation}
\frac{dE_g}{dt}=\frac{GM}{R}\frac{dm_{imp}}{dt}=\frac{16}{3}\pi^2G\bar{\rho}^2R^4\dot{R}.
\end{equation}
The energy delivered by an impact can either be retained with in the planetesimal or radiated away from the surface.
Classical impact modeling indicates that about 20-40\% of the impactor's kinetic energy is converted into thermal energy \citep[e.g,][]{okeefe+1977}. 
Subsequent studies emphasize parameter dependence: post-shock plastic deformation may enhance heating \citep{Kurosawa2021}, whereas oblique or high-velocity impacts tend to reduce it \citep{Wakita2022}, and revised melt quantification remains broadly comparable to classical estimates \citep{manske2022}. 
These findings show that the classical 20-40\% range remains a reasonable reference, within which we adopt $f_{imp} = 20\%$ as a conservative representative value for consistency with our modeling	 \citep[e.g.][]{sramek+2012, ricard+2017}, and finally the impact heating is calculated as follows, $Q_{imp}=(f_{imp}/M)(dE_{g}/dt)$.

The depth $h$ at which the energy is deposited within the icy planetesimal is related to the radius increment $dR$ by $h=6\times\,dR$, based on the modelling of the thermal evolution of growing Mars \citep{senshu+2002}.
In our model, the radius of the isobaric core during the impact, $r_{ic}$, is assumed to be equivalent to the impactor's radius, $r_{imp}$, because the planetesimal is sufficiently small (the ratio of $r_{ic}/r_{imp}$ will increase up to 1.44 for a larger size of impactors \citep{senshu+2002}).
Additionally, accretional growth is assumed to occur evenly over the entire planetesimal surface, resulting in isotropic growth, and thus, the planetesimal's radius increases by $r_{imp}$ ($\Delta R = r_{imp}$).
The release of gravitational potential energy during interior differentiation is a potential heat source that is neglected here because of the small gravity.

Although the same formulation of Qimp is used for all simulations, the resulting temperature rise varies among cases because the temporal rate of gravitational energy release ($dE_{g}/dt$), the depth of energy deposition ($h$), and local physical properties such as the water-to-rock ratio, porosity, specific heat, thermal conductivity, and pre-impact temperature control how the deposited energy is distributed between sensible heating and latent heat of melting or dehydration.
%
\subsection{Aqueous alteration}
\label{sec_aqueous_alteration}
Antigorite ((Mg, Fe)$_{3}$Si$_{2}$O$_{5}$(OH)$_{4}$), saponite ((Ca/2, Na)$_{0.3}$(Mg, Fe)$_{3}$(Si, Al)$_{4}$O$_{10}$(OH)$_{2}\cdot$\,nH$_{2}$O), and magnetite (Mg$_{3}$O$_{4}$), which have been found in CI chondrites in the form of a phyllosilicates, are the main products of aqueous alteration \citep[e.g.][]{tomeoka+1988}.
Here, we take into account only the following chemical reaction equation for aqueous alteration:\par\nointerlineskip
%
%
\begin{align}
\mathrm{\left(Mg_{0.5},\,Fe_{0.5}\right)_2SiO_4 + H_2O_{aq}} \rightarrow & \mathrm{1/3\left(Mg_{0.8},\,Fe_{0.2}\right)_3Si_2O_5\left(OH\right)_4} \nonumber \\
& + \mathrm{1/12\left(Mg_{0.8},\,Fe_{0.2}\right)_3Si_4O_{10}\left(OH\right)_2} \nonumber \\
& + \mathrm{1/4Fe_3O_4 + 1/4H_2}
\label{eq_aqueous_alteration_2}
\end{align}
We make the assumption that the values of the hydrous minerals are represented by serpentine (antigorite), for simplicity.
The timescale for aqueous alteration is known to depend strongly on temperature.
For instance, \citet{jones+2006} demonstrated through experiments that aqueous alteration completes in 15 days at 473\,K and in 60 days at 423\,K. 
Their exponential fitting suggests that, at 298\,K, the process would require between 100 and 1000 years in carbonaceous chondrites (CCs).
Similarly, \citet{wakita+2011} estimated that aqueous alteration would take approximately 210 years at 273\,K.
Given that these timescales are extremely short compared to the thermal evolution of icy planetesimals, this study assumes that aqueous alteration occurs instantaneously once ice melts.

Melting point of ice\,Ih T$_{melt}$ is given by
\begin{equation}
T_{melt}=273.0 - (1.047\times10^{-7})P
\end{equation}
where $P$ is pressure, and no any high-pressure phase ices are considered because the central pressure at the maximum radius of the planetesimals in this study, set at 1,000\,km, is approximately less than 180 MPa.

Upon the completion of aqueous alteration, the mixture of olivine and ice transforms into a mixture of hydrous minerals and excess liquid water.
As the temperature continues to rise, the ice melts, generating liquid water, causing the mass fractions of ice ($\Delta f_{mi}$) and water ($\Delta f_{mw}$) to change as follows,
\begin{equation}
\Delta f_{mw} = - \Delta f_{mi} = \frac{\overline{C_p}\left(T-T_{melt}\right)}{L_{melt}}
\end{equation}
where $L_{melt}$ is the latent heat of ice of 334 kJ/kg \citep{legates2005}.
When all the ice has melted ($f_{mi}$\,=\,0, $f_{mw}$\,=\,1), it is assumed that the hydrous minerals instantaneously migrate toward the center to form a hydrous mineral core, while the liquid water moves above the core, forming a liquid water mantle.
The liquid water layer formed by internal melting is assumed to be isothermal at the melting point of ice. 
This assumption is physically justified because the viscosity of liquid water is much lower than that of solid ice or rock, resulting in a very high Rayleigh number for the liquid layer. 
Under such conditions, vigorous thermal convection develops and rapidly redistributes heat. 
As a result, any temperature gradients within the liquid region are efficiently eliminated, allowing the layer to maintain an almost uniform temperature.
If near-surface ice melting occurs within the icy shell due to impact heating, the resulting melt is assumed to remain in place without differentiation, and its temperature is fixed at 373~K as a prescribed isothermal condition in the numerical model to avoid unrealistically high temperatures under low-pressure conditions.

As hydrous minerals in the hydrated core are further heated, dehydration reactions occur, in which hydroxyl groups are released, as represented by the following equation:\par\nointerlineskip
\begin{align}
1/3\mathrm{\left(Mg_{0.8}, Fe_{0.2}\right)_3Si_2O_5\left(OH\right)_4} &+ 1/12\mathrm{\left(Mg_{0.8}, Fe_{0.2}\right)_3Si_4O_{10}\left(OH\right)_2}\notag\\
&\rightarrow 5/8\mathrm{\left(Mg_{0.8}, Fe_{0.2}\right)_2SiO_4 + 3/4H_2O + 3/8SiO_2}. \label{eq_dehyd}\end{align}
The dehydration reaction is endothermic, with a latent heat of 417\,kJ/kg \citep{wakita+2011}.
The temperature at which dehydration occurs ($T_{dehyd}$) is given by the following relation \citep{ulmer+1995}.
\begin{equation}
T_{dehyd}=773+1.1\times10^{-7}\times P_{dehyd}
\end{equation}
Upon completion of the dehydration reaction, it is assumed that the dehydrated minerals (olivine) instantaneously migrate to the center of the planetesimal to form a dehydrated mineral core, and that the water released by the dehydration reaction is added to the water mantle.
If the temperature decreases after the dehydration reaction, aqueous alteration may occur again.
However, for simplicity, this reaction is not considered in this study.

As the temperature decreases, the water mantle solidifies to form a solid ice shell.
The thickness increase of the ice shell $\Delta r_{ice}$ over a time step $\Delta t$ is given by the following equation:
\begin{equation}
\Delta r_{ice} = \frac{F_{out} - F_{in}}{\rho_{water}L_{melt}}\Delta t ,
\end{equation}
where, $F_{in}$ is incoming heat flux from the hydrous minerals core and $F_{out}$ is outgoing heat flux.

The resulting radial heat transfer equation was solved numerically by the finite difference method using classical explicit approximations. 
The time step is dynamically adjusted according to the local thermal diffusion timescale, with an upper bound of 1 year per step to ensure numerical stability of the explicit scheme. 
Because the governing equation is of diffusion type, the explicit formulation is inherently stable under this timestep restriction and is adequate to capture the thermal evolution over the relevant Myr timescales. 
The planetesimal is represented by a one-dimensional spherical grid under spherical symmetry. 
Before the onset of aqueous alteration, the body is treated as a homogeneous ice-rock mixture; in this phase the total number of grid elements is kept constant (typically 1,000) and the grid spacing expands in proportion to the growing radius. 
Once aqueous alteration commences and compositional differentiation develops, new grid elements are added at the surface during accretion to represent material addition, and distinct grid intervals are assigned to hydrated rock, dehydrated rock, ice, and water; these intervals are recomputed at every time step. 
Spatial resolution follows the body size and characteristic heat-transport length scales, with maximum grid intervals between 1\,m (small bodies) and 1\,km (large bodies). 
This adaptive discretization maintains the resolution of phase boundaries and moving interfaces throughout the simulation.
%
\begin{table}
\begin{center}
\caption{The values of density $\rho$,\,specific heat $C_p$ and thermal conductivity $k_c$ of ice, water, rocks\,(olivine, dehydrated rocks) and hydrated rocks\,(antigorite).}
\label{table_den_speh_cond}
\begin{tabular}{lccc}
\hline
 & $\rho$ [kg\,m$^{-3}$] & $C_{p}$ [J\,kg$^{-1}$K$^{-1}$] & $k_{c}$ [W\,m$^{-1}$K$^{-1}$]\\
\hline \hline
Ice & 1000$^{}$ & 1900$^{}$ & 2.2$^{}$\\
water & 1000 & 4200 & 0.56\\
rocks& 3300$^{*}$ & 910$^{*}$ & 3.0$^{*}$\\
hydrated rocks & 2765$^{\dag}$ & 1000$^{\ddag}$ & 2.7$^{\ddag}$\\
\hline
\end{tabular}
\end{center}
\vspace{.3cm}
$*$\,\citet{wakita+2011}, $\dag$\,\citet{hilairet+2006}, $\ddag$\,\citet{osako+2010}
\end{table}
\begin{table}
\begin{center}
\caption{The values of thermal expansion coefficients and viscosities.}
\label{table_values_properties}
\begin{tabular}{lcccc}
\hline
Quantity & Symbol & Unit & Value & Reference\\
\hline \hline
Thermal expansion coefficient of ice & $\alpha_{ice}$ & K$^{-1}$ & 1.1$\times$10$^{-4}$ & \citet{roquet+2022}\\
Thermal expansion coefficient of hydrous rock & $\alpha_{hyd}$ & K$^{-1}$ & 3.9$\times$10$^{-5}$ & \citet{yang+2014}\\
Thermal expansion coefficient of dehydrated rock & $\alpha_{dehyd}$ & K$^{-1}$ & 2.4$\times$10$^{-5}$ & \citet{kirk+1987}\\
Reference viscosity of ice & $\eta_{ice}^{ref}$ & Pa\,s & $10^{15}$ & \citet{kirk+1987}\\
Reference viscosity of hydrated rock & $\eta_{serp}^{ref}$ & Pa\,s & 1.5$\times$10$^{20}$ & \citet{hilairet+2007}\\
Reference viscosity of olivine & $\eta_{oliv}^{ref}$ & Pa\,s & 4.9$\times$10$^{8}$ & \citet{karato+1986}\\
\hline
\end{tabular}
\end{center}
\end{table}
\begin{table}
\begin{center}
\caption{The values of Short-Lived Radioactive isotopes\,($^{26}$Al and $^{60}$Fe).}
\label{properties of radionuclide}
\begin{tabular}{llllclc}
\hline
Quantity & Symbol & Unit & $^{26}$Al & Ref. & $^{60}$Fe & Ref.\\
\hline \hline
Heat production per atom & E & MeV & 3.12 & 1 & 2.563 & 2,\,3\\
Decay constant & $\lambda$ & year$^{-1}$ & 9.83$\times$10$^{-8}$ & 4  & 2.65$\times$10$^{-7}$ & 5\\
Abundance in Ivuna & f & mg/g & 7967 & 6 & 182348 & 6\\
Isotope ratio at CAI formation & $^{\circ\circ}$A/$^{\times\times}$A$^{\ast}$ & N/A & 5.25$\times$10$^{-5}$ & 7 & 2.10$\times$10$^{-7}$ & 8\\
\hline
\end{tabular}
\end{center}
\vspace{.3cm}
(1)\,\citet{castillo-rogez+2009}; (2)\,\citet{ostdiek+2015}; (3)\,\citet{browne+2013}; (4)\,\citet{norris+1983}; (5)\,\citet{rugel+2009}; (6)\,\citet{braukmuller+2018}; (7)\,\citet{kita+2013}; (8)\,\citet{kodolanyi+2022}\\
*: $^{\circ\circ}$A/$^{\times\times}$A is  $^{26}$Al/$^{27}$Al for Al and $^{60}$Fe/$^{56}$Fe for Fe.
\end{table}
%
\section{Results}

\begin{figure}[p]
\begin{center}
\includegraphics[width=0.8\textwidth]{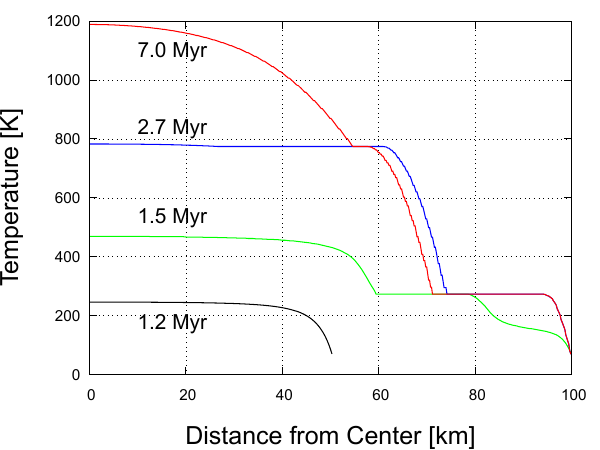} 
\end{center}
\caption{Temperature profiles in the icy planetesimal in case that accretional growth begins from 1.0\,Myr after CAI formation and accretional duration of 0.4\,Myr to a final radius of 100\,km (case\#051). 
Numbers along lines indicate times after CAI formation.
}
\label{fig_result_typical}
\end{figure}

To quantitatively investigate the thermal evolution of icy planetesimals that accrete and grow from an initial radius of 1\,km and an initial temperature of 70\,K, we conducted a series of calculations with various combinations of final radius, onset time of accretion, duration of accretion, and growth mode (Table\,\ref{table_parameter_study}).
We first present the results of a representative model (case \#051). 
Figure \ref{fig_result_typical} shows the temperature profile and its temporal evolution for the case that planetesimal growth starts at 1.0\,Myr after CAI formation from an initial radius of 1\,km to a final radius of 100\,km with $\beta=0$ for 0.4\,Myr.
Initially composed of a mixture of olivine and ice, the planetesimal increases its radius through accretion while its internal temperature rises due to radiogenic and impact heating. 
As the temperature increases and reaches the melting point of ice, melting begins at $\sim$0.2~Myr after the start of the calculation and completes on a very short timescale (less than 0.1~Myr).
The hydrous minerals settle toward the center, forming a hydrous mineral core, while excess liquid water migrates upward, creating a water mantle.
At 1.5\,Myr, the hydrous mineral core has a radius of approximately 60\,km, and the region from 60\,km to 80\,km shows that hydrous mineral and liquid water coexist and are undergoing aqueous alteration.
The region outside 80 km remains in its initial composition of a mixture of solid ice and rock and is undergoing the solid-state convection.
The hydrous mineral core continues to heat due to radioactive decay and eventually reaches the dehydration temperature.
Because dehydration reactions are endothermic, the temperature rise temporarily halts until the dehydration process is completed around 2.7\,Myr.
After completion of dehydration, a portion of the hydrous minerals transforms decreases into denser dehydrated minerals.
Consequently, the core composed of hydrous and dehydrated minerals contracts, while the thickness of the outer hydrosphere encompassing the water mantle and the ice shell increases.
This increase implicitly assumes that water released during dehydration can optimally migrate outward. 
The efficiency of such transport is uncertain, as large-scale percolation may be limited in small planetesimals \citep[e.g.][]{bland2009}, but localized fracture flow or compaction-driven expulsion could provide transient pathways. 
In this study we consider the end-member case where such transport is most efficient.

Subsequently, around 7.0\,Myr, the central temperature peaks at 1180\,K, and deeper region within 55 km consists of the dehydrate mineral, and intermediate layer between 60 and 70 km is hydrous mineral.
After that, the temperature begins to decrease as the radioactive isotopes deplete over time.
The water mantle can be retained even after 100\,Myr.
%
\subsection{Final radii}
\begin{figure}[p]
\begin{center}
\includegraphics[width=1.0\textwidth]{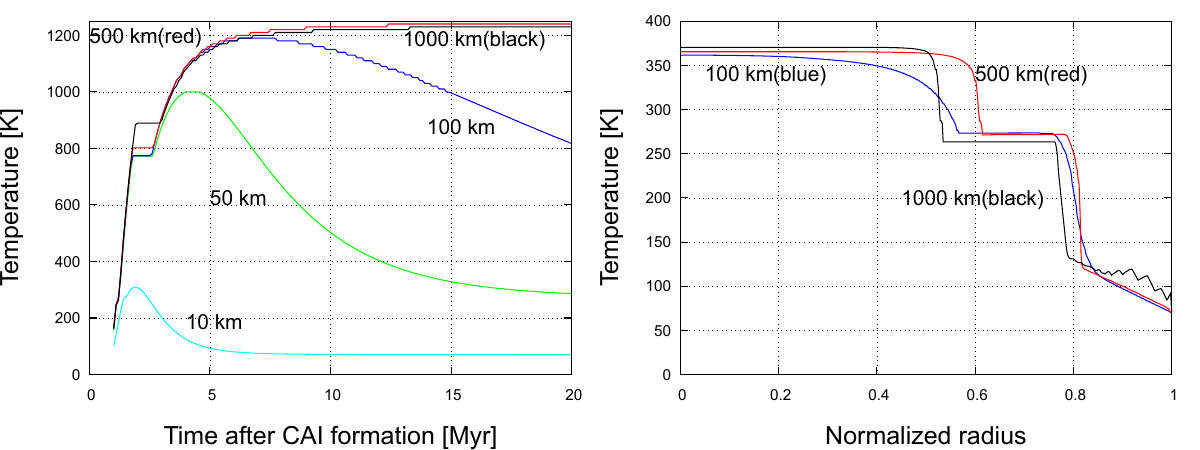} 
\end{center}
\caption{(Left) Temporal change of the central temperatures for different final radius of planetesimal, and (Right) internal temperature profiles at the end of accretional growth (at 1.4\,Myr after CAI formation) with the distance from the center normalized by the final radius of planetesimal for same calculation as shown in Figure\,\ref{fig_result_typical} (case\#051).
}
\label{fig_051}
\end{figure}

Calculations are performed assuming an growth onset 1.0\,Myr after CAI formation, an accretion duration of 0.4 Myr, and a growth mode of $\beta=0$ (case \#051), with final radii of 10, 50, 100, 500, and 1000\,km.
Figure \ref{fig_051} shows the time evolution of central temperature for each final radius, and the temperature profiles at the end of planetesimal growth.
For the final radius of 10 km, temperature increments is small and aqueous alteration occurs only in limited regions.
In case of the final radius larger than 10\,km, ice melting and aqueous alteration occur. 
If the final radius exceeds 35\,km, additional regions experiencing dehydration reactions emerge due to further temperature increases.
Larger final radii lead to broader regions of aqueous alteration and earlier onset times because the amount of radioactive isotopes increases proportionally with the mass of the planetesimal, and the melting point of ice Ih decreases at higher pressures.
Ice melting proceeds over 1$\sim$2~Myr, which is reflected in the slight kink in the central-temperature change shown in
The case of the final radii of 1000\,km shows that the post-dehydration central temperature is lower than the case of 500\,km, because the dehydration temperature increases with pressure, delaying the onset of dehydration reactions in larger bodies, during which the quantity of radioactive isotopes decreases.
In bodies with final radii of 500 km or more, the central temperature does not decrease significantly even after a few tens of Myr, because their smaller surface-to-volume ratio and the extensive dehydration limit cooling to slow conduction, making heat loss inefficient.
Although only short-lived radioactive isotopes are considered as heat sources in this study, the inclusion of long-lived isotopes would further slow the cooling process.
Therefore, larger final radii result in higher internal temperatures and longer durations of elevated thermal states.
The temperature profiles at the end of planetesimal growth reveals that near surface regions experience heating due to impact energy, with the effect becoming larger for a larger final radius.
The jagged profile at the surficial region for the final radius of 1000\,km (Figure \ref{fig_051}, right) are due to simultaneous radius growth and impact heating.
While the imprint of impact heating becomes more apparent at larger final radii, this reflects the increase in accretional impact energy as bodies grow: larger bodies intercept more impactors, and faster mass growth brings more frequent and higher-velocity collisions. 
The heated layer nevertheless remains confined to the outermost $\sim$10\% of the radius and does not drive global aqueous alteration or dehydration.
%
\subsection{Growth mode}
\label{sec_growthmode}
\begin{figure}[p]
\begin{center}
\includegraphics[width=1.0\textwidth]{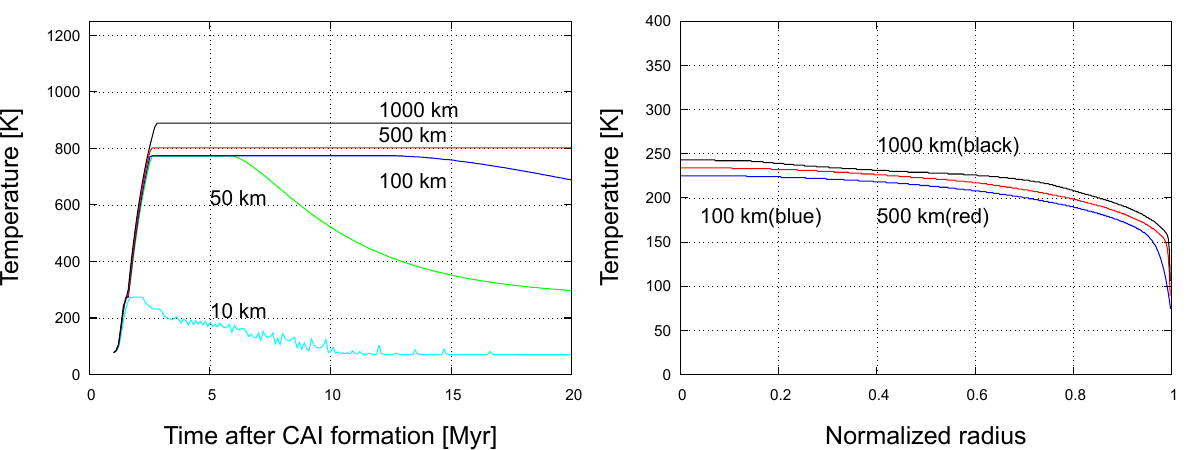} 
\end{center}
\caption{Same as Figure\,\ref{fig_051} but with the growth mode of $\beta=2$ (case\#061).}
\label{fig_061}
\end{figure}

Temperature evolution at the center for the runaway growth mode of $\beta=2$ (case \#061) is shown in Figure \ref{fig_061}.
Compared to the linear growth mode of $\beta=0$ (Figure \ref{fig_051}), for the final radius of 10\,km, the central region reaches the melting point of ice and undergoes aqueous alteration. 
However, the ice does not fully melt, and the temperature eventually decreases.
For the final radius of 50\,km or larger, ice melting and aqueous alteration occur in some regions, but further temperature increases are limited. 
Although the central region reaches the dehydration temperature, the dehydration reaction does not complete, and the region eventually begins to cool.
Such results are attributed to a large depletion of radioactive isotopes before the planetesimal undergoes rapid growth at the final accretion stages ($\beta=2$).
Temperature profiles with final radii of 100, 500, and 1000 km at the end of planetesimal growth (at 1.4\,Myr after CAI formation) show lower temperatures in the deeper regions compared to the previous case of $\beta=0$, due to the reasons discussed above, and higher (up to 100\,K) temperatures near the surface as a result of impact heating due to rapid growth.
The earlier the growth of a planetesimal, the more radioactive isotopes it retains, leading to higher internal temperatures.
%
\subsection{Starting period and duration of planetesimal growth}
\begin{figure}[p]
\begin{center}
\includegraphics[width=1.0\textwidth]{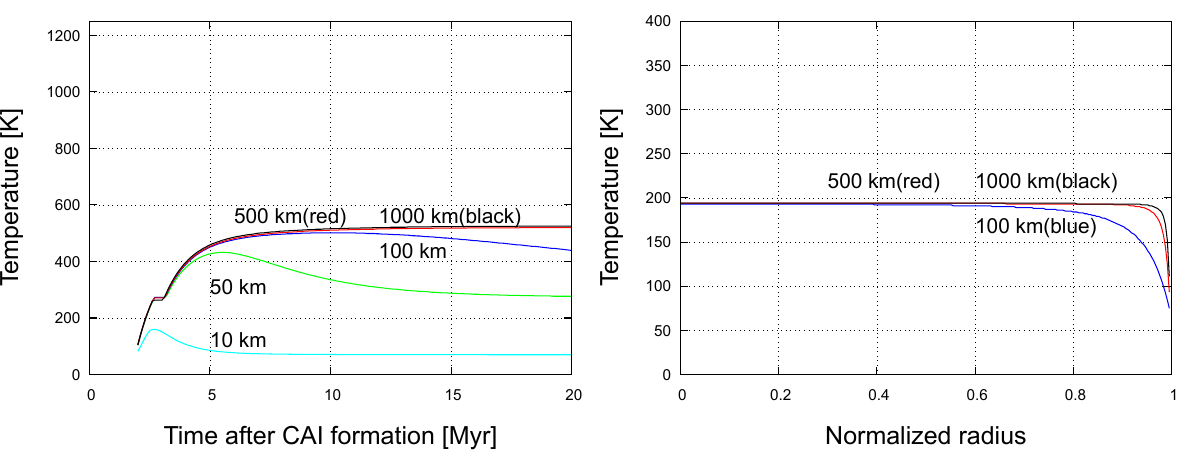} 
\end{center}
\caption{Same as Figure\,\ref{fig_051} but the start timing of the accretional growth was changed from 1.0\,Myr to 2.0\,Myr after the formation of the CAI (case\#054).
Note that the internal temperature profiles (right) are at the end of accretion (at 2.4\,Myr after CAI formation).}
\label{fig_054}
\end{figure}

\begin{figure}[p]
\begin{center}
\includegraphics[width=1.0\textwidth]{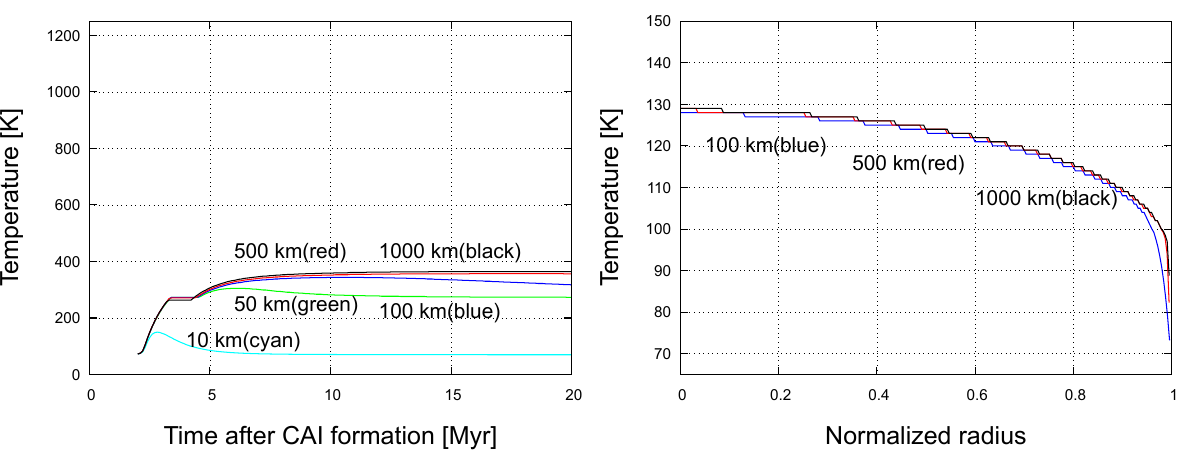} 
\end{center}
\caption{Same as Figure\,\ref{fig_054} but with the growth mode of $\beta=2$ (case\#064).}
\label{fig_064}
\end{figure}

Because short-lived radioactive isotopes, which serve as the primary heat sources during planetesimal growth, have half-lives of approximately 1\,Myr, the timing of their growth strongly influences the initial amount of heat retained within the planetesimal.
Thermal evolution for a case where planetesimal growth begins 2.0\,Myr after CAI formation is shown in Figure \ref{fig_054} (case\#054).
Compared to the case with growth starting at 1.0\,Myr (case\#051, Figure \ref{fig_051}), the peak central temperature is significantly lower.
In the case with a final radius of 10\,km, no ice melting occurs in any region of the body, and consequently, no aqueous alteration takes place.
For final radii of 50\,km or more, partial ice melting and aqueous alteration are observed in some regions, although the temperature never reaches the dehydration point.
In the case of runaway growth ($\beta=2$, case\#064), the increase in radius occurs later, resulting in even lower peak temperatures (Figure \ref{fig_064}).
As in the previous case, no melting occurs when the final radius is 10 km.
For bodies with final radii greater than 50\,km, ice melting and aqueous alteration are observed in some regions; however, peak temperatures are approximately 200 K lower than the previous case.
These findings suggest that delayed growth leads to only limited temperature increases compared with cases of earlier accretion, primarily because most radioactive isotopes have already decayed before the planetesimal reaches a substantial size.

\begin{figure}[p]
\begin{center}
\includegraphics[width=1.0\textwidth]{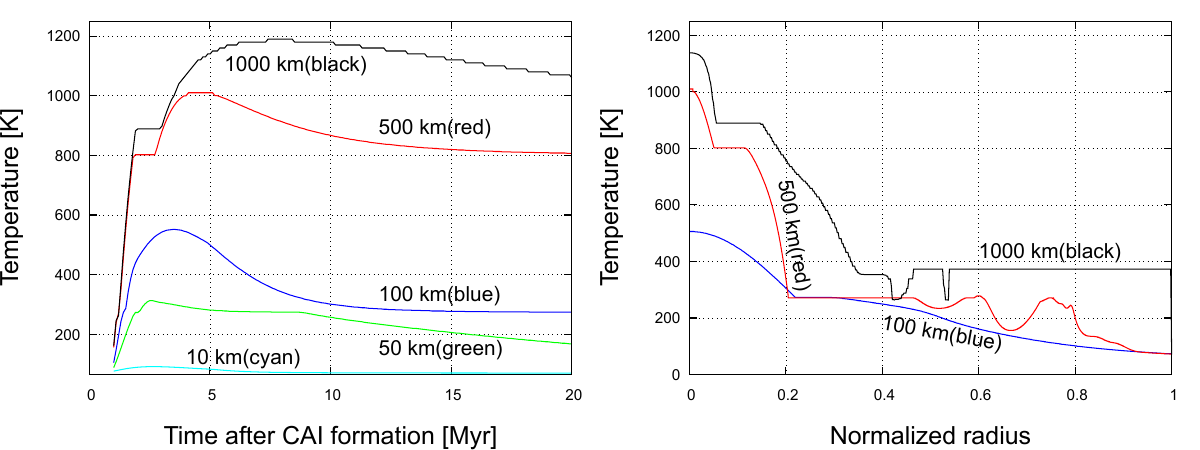} 
\end{center}
\caption{Same as Figure\,\ref{fig_051} but the duration of the accretional growth is changed from 0.4\,Myr to 4.0\,Myr (case\#052).
Note that the internal temperature profiles (right) are at the end of accretion (at 5.0\,Myr after CAI formation).}
\label{fig_052}
\end{figure}

Figure \ref{fig_052} shows the thermal evolution for a growth duration of 4.0\,Myr (case\#052).
Compared to the case with the growth duration of 0.4\,Myr (Figure \ref{fig_051}), there is almost no temperature change from the initial condition when the final radius is 10\,km.
For bodies with final radii 50\,km or more, ice melting and aqueous alteration occur, and for radii of 500 km or more, dehydration reactions are completed in some regions, resulting in the formation of a dehydrated mineral core.
Immediately after growth is complete, the temperature profile exceeds the dehydration point at the center, while the temperature in the outer region remains relatively low.
In the 500 km case, a distinct temperature peak caused by impact heating appears, and in the 1000\,km case, more than half of the outer region experiences significant heating due to impacts (Figure \ref{fig_052}).

\begin{figure}[p]
\begin{center}
\includegraphics[width=1.0\textwidth]{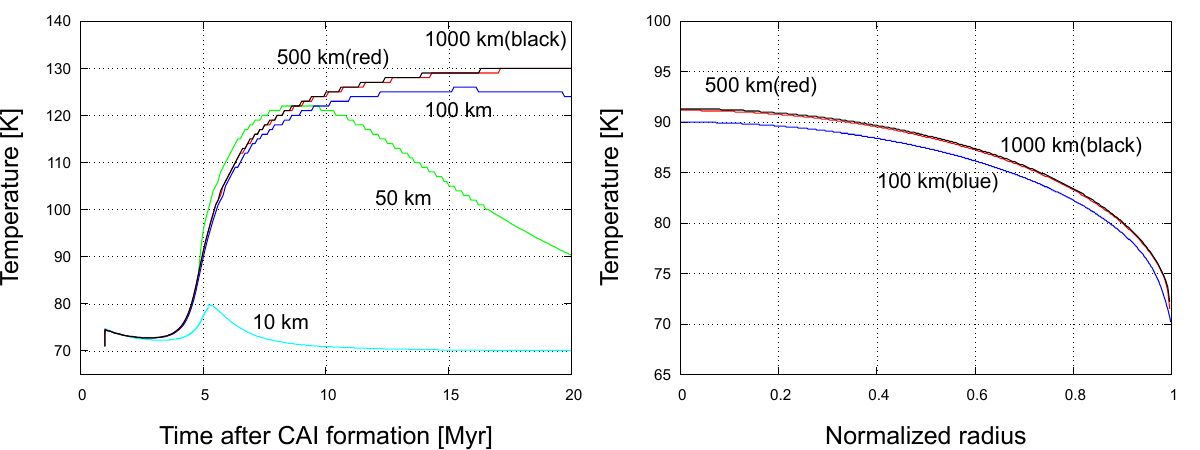} 
\end{center}
\caption{Same as Figure\,\ref{fig_052} but with the growth mode of $\beta=2$ (case\#062).}
\label{fig_062}
\end{figure}

In the case of growth mode $\beta=2$ (case\#062, Figure \ref{fig_062}), the planetesimal radius increases after the depletion of radioactive isotopes, resulting in no temperature increase compared to the cases where growth is completed at 1.4\,Myr (Figure \ref{fig_061}, case\#061) and 2.4\,Myr (Figure \ref{fig_064}, case\#064) after CAI formation.

\begin{figure}[p]
\begin{center}
\includegraphics[width=1.0\textwidth]{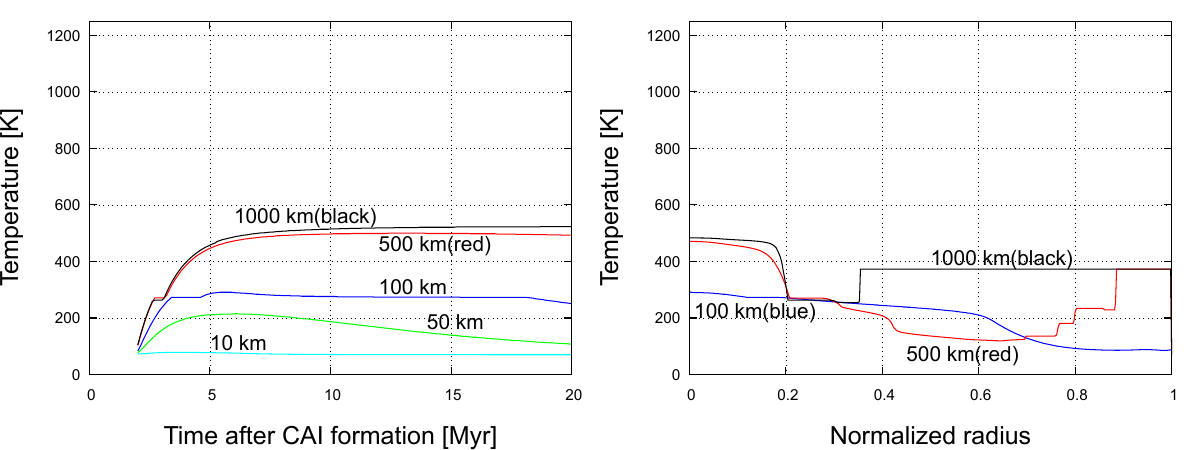} 
\end{center}
\caption{Same as Figure\,\ref{fig_051} but but the duration of the accretional growth is changed from 0.4\,Myr to 4.0\,Myr (case\#055).
Note that the internal temperature profiles (right) are at the end of accretion (at 6.0\,Myr after CAI formation).}
\label{fig_055}
\end{figure}

The results for the case in which both the growth onset time and growth duration are changed --from 1.0\,Myr to 2.0\,Myr after CAI formation, and from 0.4\,Myr to\,4.0 Myr, respectively-- are shown in Figure \ref{fig_055} (case\#055).
The central temperature evolution is similar to that in the case with a delayed growth onset (case\#054), and immediately after growth is completed, moderate heating is observed in the core, along with impact-induced heating near the surface.
This thermal evolution, characterized by limited central heating, also resembles the result of the case with a prolonged growth duration (case\#052), and is attributed to the decay of radioactive isotopes before the body reaches a large radius, as in the case of delayed onset.
In the case of linear growth ($\beta=0$), low temperatures in the outer regions are due to the addition of material after significant isotope decay.
Furthermore, the impact heating observed immediately after growth completion is thought to result from the prolonged duration of growth.
With a longer growth duration, the mass of material added per unit time is smaller, but the icy body remains massive for a longer time.
As a result, high-velocity impactors continue to collide with the body over an extended duration, enhancing the overall effect of impact heating.
%
\subsection{Maximum temperature, structure of interior and implications for actual objects}

\begin{figure}[t]
\begin{center}
\includegraphics[width=1.0\textwidth]{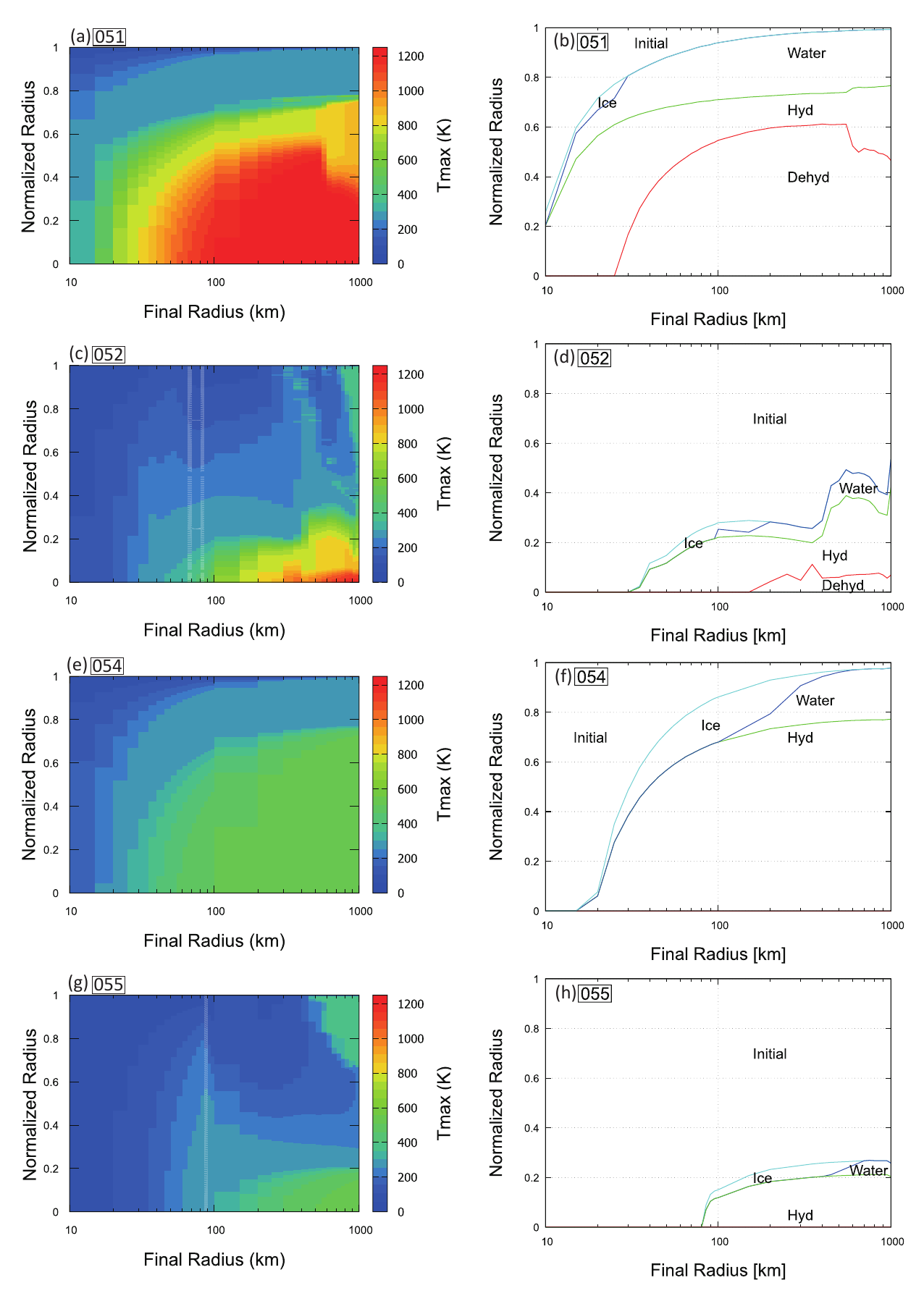} 
\end{center}
\caption{Maximum temperatures of the planetesimal interior for the cases of 051 (a), 052 (c), 054 (e) and 055 (g).
Interior structures at 100\,Myr after CAI formation for the case of 051 (b), 052 (d), 054 (f) and 055 (h). 
The vertical axes are normalized planetesimal radius by the final surface radius.}
\label{fig_max_r_0}
\end{figure}

\begin{figure}[t]
\begin{center}
\includegraphics[width=1.0\textwidth]{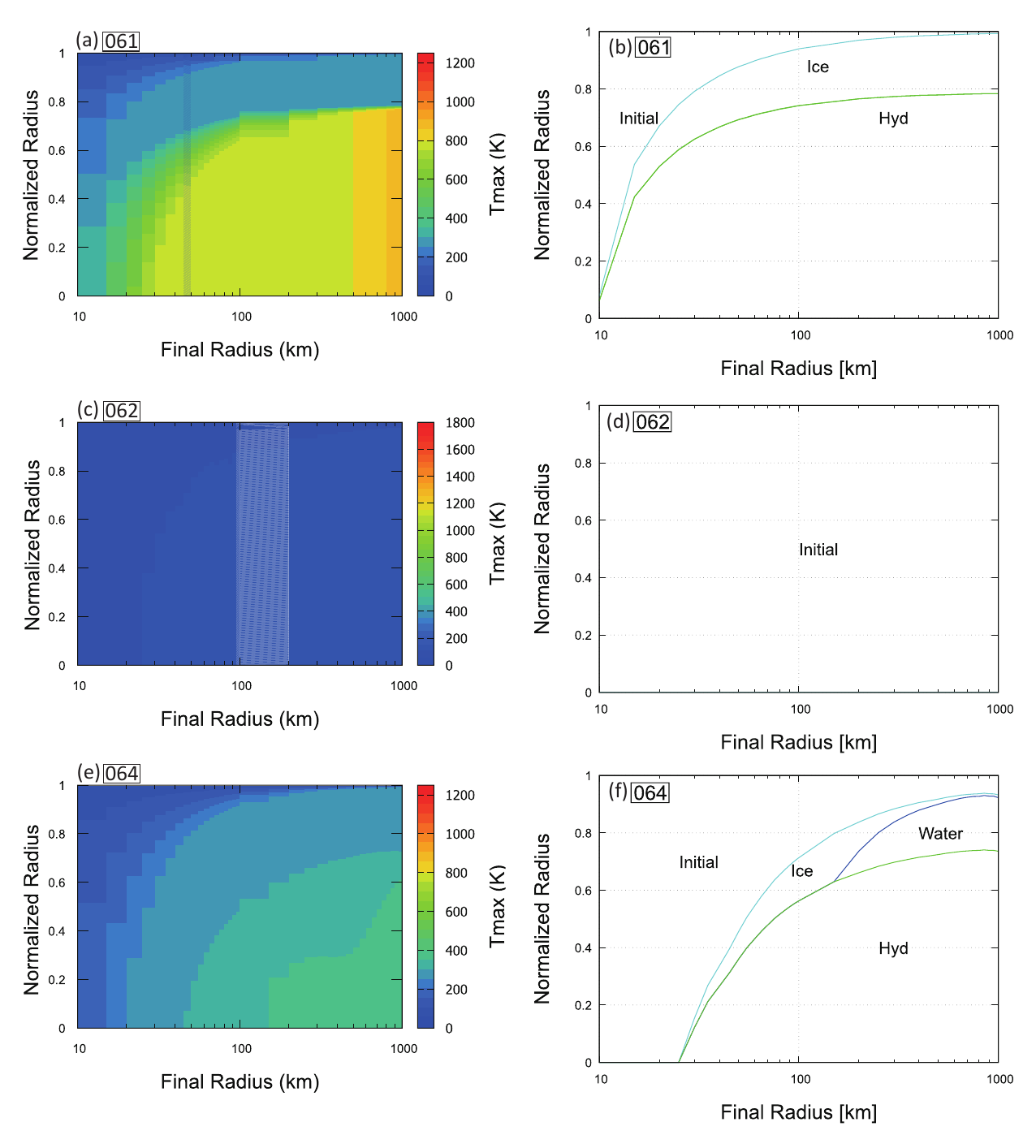} 
\end{center}
\caption{Same as Figure\,\ref{fig_max_r_0} but with the growth mode of $\beta=2$.}
\label{fig_max_r_2}
\end{figure}

To examine how variations in parameters affect the internal structural evolution of planetesimals, Figure \ref{fig_max_r_0} (growth mode of $\beta=0$) and Figure \ref{fig_max_r_2} ($\beta=2$) show the interior maximum temperatures and the interior structures at 100\,Myr after CAI formation.
In the case where planetesimal growth begins 1.0\,Myr after CAI formation and lasts for 0.4\,Myr with $\beta=0$ (case\#051, Table \ref{table_parameter_study}), the icy planetesimal with a final radius of 100\,km reaches maximum temperatures exceeding 1000\,K at depths greater than 40\,km (Figure \ref{fig_max_r_0}a).
Its interior structure consists of a dehydrated mineral core extending to a radius of approximately 55\,km, surrounded by a hydrated mineral layer from 55\,km to 70\,km, a water mantle from 70\,km to 95\,km, and an outer shell of primordial composition (a mixture of olivine and ice) from 95\,km to the surface (Figure \ref{fig_max_r_0}b).
Although this layer is denser than the underlying water/ice, its thinness, the weak gravity of a $\sim$100~km body, and the high strength and viscosity of the cold outer material can suppress significant gravitational overturn on the timescale considered here. 
Its persistence should therefore be regarded as conditional and not guaranteed under all circumstances.
In the case for a final radius of 1000\,km, the region between 450 km and 750 km experience maximum temperatures around 900\,K, while the outer layer greater than 800\,km remain at approximately 300\,K (Figure \ref{fig_max_r_0}a).
Its interior consists of a dehydrated mineral core with a radius of about 500\,km, a hydrated mineral layer between 500\,km and 770\,km, a water mantle from 770\,km to 990\,km, and an outer shell of primordial composition extending from 990\,km to the surface (Figure \ref{fig_max_r_0}b).
When the final radius is 10~km, the entire hydrated layer maintains temperatures of 300-350~K, and for final radii below~100 km this temperature range is limited to only a few kilometers near the surface of the hydrated region.

If the end of planetesimal growth is later (case\#052), the depletion of radioactive isotopes results in the preservation of the initial composition in about half of the body, even for a final radius of 1000 km (Figure \ref{fig_max_r_0}c and \ref{fig_max_r_0}d).
In this region, neither large temperature increasing nor aqueous alteration has occurred, suggesting a high likelihood that the primordial composition from the time of Solar System formation could be preserved.
In the case\#054, accretional growth begins later than in the case\#051, and the radioactive heat source is more depleted, resulting in a lower temperature increase (Figure \ref{fig_max_r_0}e and \ref{fig_max_r_0}f).
In the case\#055, the duration of accretional growth is even longer, leading to an even smaller temperature rise (Figure \ref{fig_max_r_0}g, h).
The late growth mode ($\beta=2$) limits the temperature increment (Figure \ref{fig_max_r_2}) because a large depletion of radioactive isotopes before the planetesimal undergoes rapid growth at the final stage (see, section \ref{sec_growthmode}).

Low-temperature regions that retain their initial composition without experiencing ice melting or aqueous alteration can extend deeper into the body when the final radius is smaller and the time from CAI formation to the completion of accretional growth is longer (cases\#052, \#055 and \#062).
Such a region is likely to retain its composition since the formation of the Solar System.
When the initial ice-rock mixture is heated and ice begins to melt, aqueous alteration of the rock occurs.
In case\#052, when the final radius is smaller than 70~km, the entire hydrated layer remains within the 300-350~K range. For final radii between 70 and 200~km, however, only a few kilometers near the surface of the hydrated layer stay within this temperature range. 
In case\#054, the entire hydrated layer maintains temperatures of 300-350~K when the final radius is 30 km, while for larger radii, this range is confined to a few kilometers near the surface of the hydrated region. 
In case\#055, when the final radius is 100 km, the entire hydrated layer also remains within the 300-350~K range. 
For final radii between 100 and several hundred kilometers, however, this temperature range is again limited to only a few kilometers near the surface. 
Similarly, in case\#061, the whole hydrated layer stays at 300-350~K when the final radius is 10 km, whereas for final radii between 20 and 100~km, this range is restricted to just a few kilometers near the surface. These regions may represent potential sources of Ryugu materials.
%
%

On the other hand, aliphatic hydrocarbons found in CI and CM chondrites are unstable; their C-H bonds degrade within 200\,years at 373\,K \citep{kebukawa+2010}. 
Since the temperature in the hydrous minerals region shown in Figure \ref{fig_max_r_0} and \ref{fig_max_r_2} partly exceed 373\,K, CI and CM chondrites likely formed in limited regions that experienced aqueous alteration through an incomplete ice melting (f$_{mw}>$0 and f$_{mi}<$1), e.g., in the case of a final radius of 100\,km in Figure 11e (case\#064), the region less than 50 km in radius with the temperature between 273\,K and 373\,K.

Cases where the ice has melted and a liquid water layer can exist (cases \#054 and \#064) may represent the origin of the icy body with the subsurface ocean. 
For example, Saturnian icy moon Enceladus has an average surface radius of 252.1$\pm$0.2\,km \citep{thomas+2007}, and its hydrous mineral core is estimated to be between 180 km and 200 km which is covered by the subsurface ocean of 20--40 km in thickness \citep{schubert+2006}. 
In case\#064, most parts of the hydrated layer remain at around 350~K, which may also correspond to the region from which Ryugu materials originated.

A dehydrated mineral core forms at the center experiencing strong heating, particularly when accretional growth starts earlier and the radius increases linearly ($\beta=0$). 
In such cases, the central temperature can reach up to the Fe-FeS eutectic point of 1250\,K, potentially causing metallic melting (Figure \ref{fig_max_r_0}a and \ref{fig_max_r_0}c). 
Iron meteorites with compositions similar to CCs are thought to originate from such regions \citep[e.g.][]{kruijer2017,kruijer2020}.
Note that in case\#051, when the final radius exceeds 600\,km, the radius of the dehydrated core decreases with increasing body size (Figure \ref{fig_max_r_0}d).
This is because the dehydration temperature rises with pressure, requiring more time to reach that temperature. 
During this period, radioactive heat sources decay, which limits the extent of dehydration.

\begin{figure}[t]
\begin{center}
\includegraphics[width=1.0\textwidth]{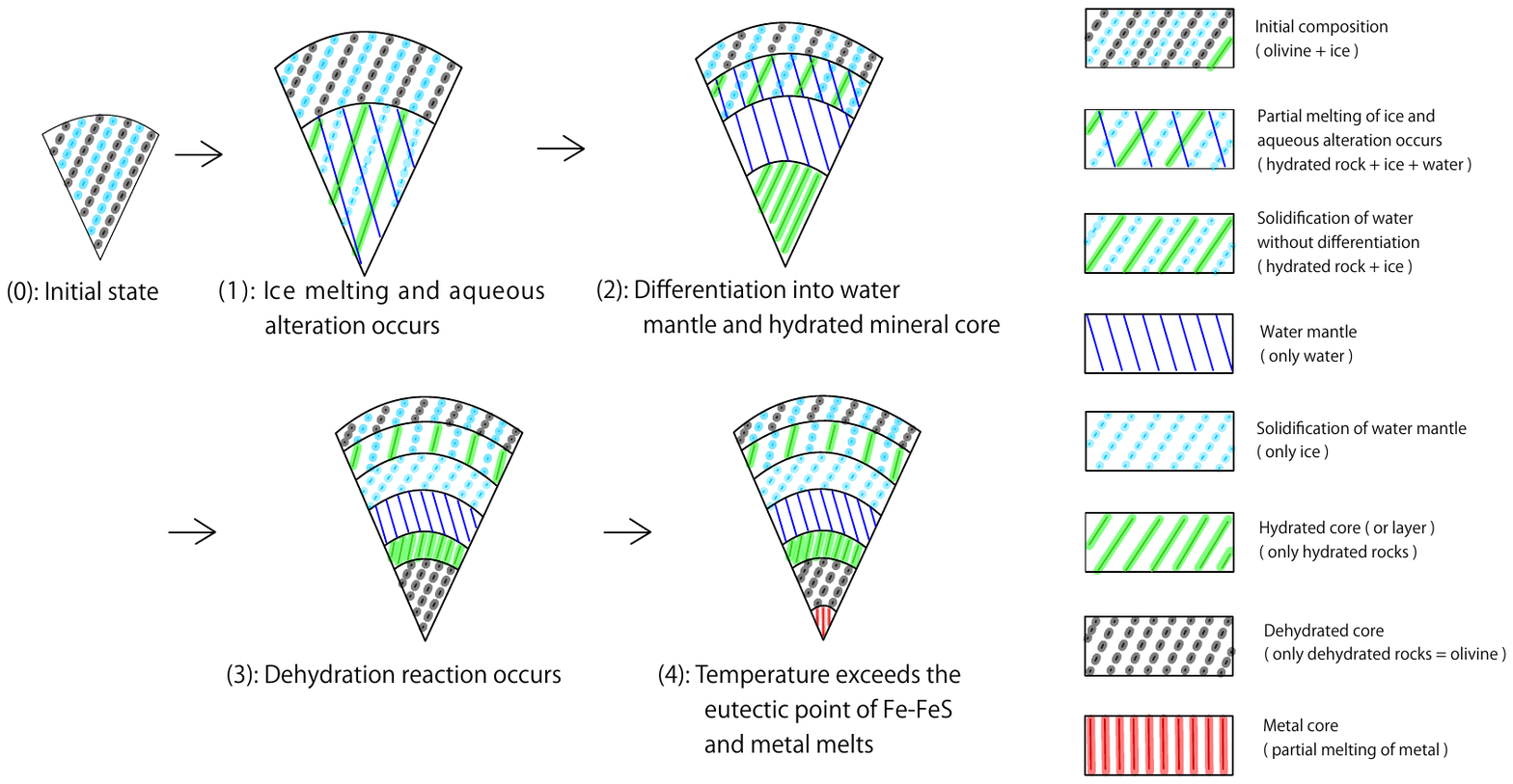}
\end{center}
\caption{Schematic diagram of interior structures and their evolution of the icy planetesimals.
Initial structure is composed of a mixture of ice and olivine (0); as temperature increases, ice melts and aqueous alteration occurs (1); later, when all ice melts, water mantle and hydrated mineral core differentiate (2); further heating of the hydrated mineral core, dehydration reaction occurs, resulting in a dehydrated mineral core (3); if temperature of the dehydrated mineral core increases and exceeds Fe--FeS eutectic point (1250\,K), a metallic core can be formed (4).}
\label{fig_result_summary}
\end{figure}

The thermal evolution and resulting internal structures of icy planetesimals derived from this study, including those 100 million years after CAI formation, are summarized in Figure \ref{fig_result_summary} and Table \ref{table_internal_structure}. 
These findings provide a unified framework for discussing the diverse thermal histories of icy planetesimals.
The evolutionary structures obtained in this study (e.g., Figure \ref{fig_result_summary}) resemble those reported by \citet{neumann2020} for water-rich planetesimals and Kuiper belt objects, particularly in the development of hydrated cores, water mantles, and dehydrated interiors. 
This similarity reflects the universality of aqueous alteration and differentiation processes in icy bodies. 
However, whereas \citet{neumann2020} focused on larger, often differentiated bodies, our work addresses smaller, undifferentiated icy planetesimals, adopting a representative W/R ratio (0.5) consistent with carbonaceous chondrite compositions, in order to place constraints on the thermal and structural evolution of Ryugu's parent body.
\clearpage
\begin{table}
\begin{center}
\caption{List of Parameters of accretion used in this study.}
\label{table_parameter_study}
\begin{tabular}{lccc}
\hline
\# & $\beta$ & Starting timing of accretion & Duration of \\
 &  & after CAI formation [Myr] & accretion [Myr] \\
\hline \hline
051 & 0 & 1.0 & 0.4 \\
052 & 0 & 1.0 & 4.0 \\
054 & 0 & 2.0 & 0.4 \\
055 & 0 & 2.0 & 4.0 \\
061 & 2 & 1.0 & 0.4 \\
062 & 2 & 1.0 & 4.0 \\
064 & 2 & 2.0 & 0.4 \\
\hline
\end{tabular}
\end{center}
\end{table}
\begin{table}
\begin{center}
\caption{Interior structures at 100\,Myr after CAI formation for different parameters and different final radii. Each number corresponds to the interior structure number shown in Figure\,\ref{fig_result_summary}.}
\label{table_internal_structure}
\begin{tabular}{cccccc}
\hline
Model\# & 10\,km & 50\,km & 100\,km & 500\,km & 1000\,km \\
\hline \hline
051 & (2) & (3) & (3) & (4) & (4) \\
052 & (0) & (2) & (2) & (3) & (4) \\
054 & (1) & (2) & (2) & (2) & (2) \\
055 & (0) & (0) & (2) & (2) & (2) \\
061 & (2) & (2) & (2) & (2) & (2) \\
062 & (0) & (0) & (0) & (0) & (0) \\
064 & (0) & (2) & (2) & (2) & (2) \\
\hline
\end{tabular}
\end{center}
\end{table}
\clearpage
\section{Discussions}
Before examining individual factors that may modify the thermal evolution of icy planetesimals, we briefly summarize the key findings from the preceding section. 
Our simulations demonstrate that the interior temperature and structure of icy planetesimals depend strongly on the timing and duration of accretion as well as the final radius. 
Bodies that began accretion around 2~Myr after CAI formation and reached final radii of several tens of kilometers maintained hydrated layers at 300-350~K, whereas earlier or larger bodies developed higher temperatures confined to their deep interiors, leaving only the shallow parts of the hydrated layer within this range. 
These results indicate that low-temperature aqueous alteration and the preservation of primitive materials could occur only in restricted regions of the hydrated layer. In the following sections, we first discuss the implications of these results for Ryugu's parent body and then examine additional processes such as porosity evolution, impact heating, and variations in the water-to-rock ratio, which could further influence the thermal and chemical evolution of icy planetesimals.

\subsection{Implications for Ryugu's parent body}
The timing and temperature of aqueous alteration inferred from Ryugu samples provide essential constraints for evaluating our numerical results. Mn-Cr chronology indicates that alteration occurred between 1.8 and 5.2 Myr after the formation of CAIs, while carbonate and sulfide mineral thermometry suggests alteration temperatures of approximately 20-40~${}^\circ$C (Nakamura et al. 2022; Yokoyama et al. 2023; McCain et al. 2023). These findings imply that Ryugu's parent body experienced mild heating and limited dehydration.
Our simulations show that such low-temperature alteration conditions can be reproduced in several parameter sets. In cases \#052, \#054, \#055, and \#061, the hydrated layer maintains temperatures of 300-350~K either throughout the interior when the final radius is smaller than 30-70~km, or within a few-kilometer-thick surface zone of the hydrated layer for larger bodies up to a few hundred kilometers. In case \#064, most parts of the hydrated layer remain at around 350~K. These regions correspond to those expected to retain hydrous minerals without dehydration and therefore represent potential sources of the Ryugu materials. The temperature range of 300-350~K derived from the model agrees well with the 20-40 ${}^\circ$C inferred from sample analyses, indicating that Ryugu's parent body likely accreted relatively late, about 2 Myr after CAI formation, with a final radius of several tens of kilometers.
Earlier accretion ($<~$ 1 Myr after CAIs) or much larger final radii ($>$ 100 km) yield central temperatures exceeding 600~K, leading to extensive dehydration that is inconsistent with the observed mineralogy. 
However, even in these larger bodies, only the outer few kilometers of the hydrated layer remain within 300-350~K. 
It is therefore plausible that Ryugu materials originated from such shallow parts of the hydrated layer in a larger icy planetesimal that later experienced collisional disruption, with those low-temperature hydrous fragments reaccumulating to form Ryugu.
Moreover, the stability limit of aliphatic hydrocarbons, whose C-H bonds degrade within 200 years at 373~K (Kebukawa et al. 2010), indicates that the preservation of organic matter in Ryugu requires regions remaining below 373~K. 
The modeled 300-350~K domains therefore provide a physically consistent explanation for the coexistence of hydrous minerals and preserved organics.
Collectively, these comparisons constrain the Ryugu parent body to have accreted a few million years after CAI formation, with a radius of several tens of kilometers, maintaining low-temperature aqueous alteration in its shallow interior while retaining primitive materials near the surface, or alternatively, that Ryugu formed from the low-temperature outer part of the hydrated layer of a larger icy body disrupted by impacts.

\subsection{Effects of porosity}
Asteroid Ryugu has a rubble--pile structure, and its estimated bulk porosity is as high as 58\% \citep{grott+2020}, which includes both micro- and macro-porosity components. 
The macro-porosity mainly reflects the voids between re-accreted fragments, whereas the micro-porosity represents the fine-scale pore spaces within the constituent grains. 
According to \citet{Sakatani2021}, the upper bound of the micro-porosity of Ryugu's surface materials is considerably lower than the bulk value. 
Maintaining high porosity under the internal pressures of large bodies is problematic, and previous studies suggest that micro-porosity would be largely eliminated in the deep interior \citep[e.g.][]{neumann2021}. 
Therefore, porous materials were likely preserved only in shallower layers, which may have provided the main source of Ryugu's constituent materials.
In general, small bodies with low gravity tend to have high porosity. 
As these bodies grow and their mass and gravity increase, porosity decreases owing to compaction and sintering induced by heating.
High porosity has conflicting effects on thermal evolution. 
Thermal conductivity would be reduced, making it easier for the body to heat up. 
On the other hand, the amount of radioactive heat sources would be lower and fluids would be easer to move through the pores, enhancing heat transport and making it harder for the body to warm up.
\citet{yomogida+1984} conducted numerical calculations on the thermal evolution of ordinary chondrite parent bodies, incorporating porosity reduction from an initial value of 40\% due to sintering. 
They reported that reducing porosity increases the cooling rate.
\citet{lichtenberg+2016} investigated the thermal evolution of silicate bodies that had not undergone aqueous alteration. 
Using numerical models with some parameters; initial porosity, body radius, and formation time, they found that lower thermal conductivity due to higher porosity promotes heating. 
However, they also noted that the effect of porosity is smaller compared to that of body size and formation timing.
\citet{wakita+2011} performed a simple calculation, varying thermal conductivity by factors of 0.5 and 2.0 to examine its effects on icy planetesimals. 
They concluded that halving the porosity led to an increase in the size of the differentiated rocky core and the thickness of the surrounding water mantle, but had little effect on the timing and duration of ice melting.
Although \citet{golabek+2021} also considered the porosity in their numerical model, their model did not include temporal change and parameter studies of the porosity.
Quantitatively assessing the effects of porosity is important for understanding the thermal evolution of icy planetesimals, and future studies are expected to incorporate it into the model.

Recent models have explicitly treated the coupled thermal and porosity evolution of hydrated and icy bodies. 
\citet{neumann2015, neumann2021, neumann2024} showed that compaction by creep and hot pressing largely controls the loss of porosity during aqueous alteration, from Ceres-sized bodies to the parent body of Ryugu. 
Their results indicate that small ($<$10~km) bodies can remain highly porous, whereas larger ones lose porosity except in their outer shells. 
\citet{neumann2019} demonstrated that Enceladus' partially compacted core can retain a porous outer layer that sustains hydrothermal circulation. 
\citet{malamud2013, malamud2015} obtained similar implications for Enceladus and Kuiper Belt objects, where porosity evolution regulates both conductive and advective heat transport. 
\citet{bierson2019} further showed that radiogenic heating can cause porosity collapse in large KBOs once internal temperatures approach the ice melting point. 
Collectively, these studies demonstrate that porosity co-evolves with temperature and permeability, coupling body size and accretion timing to the extent of aqueous alteration. 
Our present model focuses on the low-mass end of this range, that is, small and undifferentiated icy planetesimals, where porosity exerts a key influence on thermal evolution and will be incorporated more completely in future work.

\subsection{Effect of impact velocity on accretional heating}
In this study, we assumed that the impact velocity $v$ of accreting bodies is equal to the escape velocity $v_{eq}\propto M^{1/2}$. 
On the other hand, accreting bodies also exhibit random velocities \citep[cf.][]{stewart+1988, wetherill+1989}.
These eccentricities and inclinations follow a normal distribution \citep{ida+1992}. 
When the optical depth of the protoplanetary disk is low ($\sim$ 0.3), the radial component of the random velocity also follows a normal distribution \citep{daisaka+1999, ohtsuki+2000}. In the following discussion, we define $v_{ran}$ as the mean of the velocity distribution governed by a normal distribution.
During runaway growth, planetary embryos of mass $M$ and surrounding planetesimals approach equilibrium through dynamical friction, leading to a scaling of $v_{ran}\propto M^{-1/2}$.
Later, gravitational perturbations from these embryos stir the orbits of nearby planetesimals, and $v_{ran}$ evolves depending on both viscous heating and gas drag. When these effects reach equilibrium, $v_{ran}\propto M^{1/3}$ \citep{ida+1993, sasaki+2019}. 
Therefore, the impact velocity used in this study may be overestimated compared to actual accretion conditions.
However, while assuming random velocities would result in lower surface temperatures, the qualitative implications of our results regarding impact heating are not expected to change.

\subsection{Water rock mass ratio of icy planetesimals}
Current lack of direct exploration of icy planetesimals leads that their W/R remains highly uncertain. 
For example, estimates range from W/R\,=\,1.72 \citep{anders+1989, wakita+2011} to 0.2--0.9 \citep{nakamuraT+2023}.
This uncertainty significantly affects thermal evolution, as the amount of radiogenic heat is directly tied to the rock fraction.
Here, we assume W/R\,=\,0.5. 
In this study, the W/R for the initial composition of the planetesimal is set to 0.5, as described in Section 2.2. 
This corresponds to approximate volume fractions of $\sim$62\% ice and $\sim$38\% dry silicates before hydration. After hydration, part of the water is incorporated into hydrated silicates and the remainder remains as free water. 
For W/R = 0.5, this yields $\sim$54\% hydrated silicates and $\sim$46\% water by volume, with updated mass fractions of $\sim$77\% hydrated silicates and $\sim$23\% water.
In contrast, \citet{wakita+2011} used a value of W/R\,=\,2.0 and conducted calculations without considering thermal convection or progressive accretion.
Time evolution of planetesimal radius in our model assuming the growth mode of $\beta=2$ is similar to the instantaneous accretion assumed by \citet{wakita+2011}. 
When comparing our case\#061, where accretion completes 1.4\,Myr after CAI formation, we find that for a final radius of 10\,km, the temperature rise is minimal. 
However, for radii above 50\,km, the temperature exceeds the dehydration point, after which rapid cooling occurs.
Such difference arises because \citet{wakita+2011} assumed a lower rock content in the initial composition, resulting in less radiogenic heat.
Therefore, when W/R$<$1.0, higher peak temperatures are expected. 
Conversely, when W/R$>$1.0, the lower rock content reduces internal heating, and enhanced thermal convection leads a smaller temperature increase.
However, once the temperature reaches the melting point of ice, the abundant water likely promotes aqueous alteration.

\subsection{Future tests and spacecraft exploration for icy planetesimals}
Our thermal evolution models suggest that icy planetesimals can preserve hydrated interiors or develop differentiated structures depending on their accretion histories. 
D-type asteroids, which are considered remnants of such icy planetesimals, may thus exhibit internal structures consistent with those derived from our models.
D--type asteroids, which are thought to be rich in organics and water, exist farther from the Sun than the region where C--type asteroids, such as Ryugu, are commonly found. 
These D--type asteroids are considered to be similar in nature to icy planetesimals. 
Most of the Jupiter Trojan asteroids are classified as D--type \citep[cf.][]{carvano+2010} and are thought to be remnants of material that failed to become part of Jupiter's icy satellites.
NASA launched the Lucy spacecraft in 2021 to conduct close exploration of the Jupiter Trojans.
The spacecraft is scheduled to arrive in the Trojan region in 2027 and will perform flybys of multiple Trojan asteroids until around 2033. 
Lucy is equipped with a visible imager, L'LORRI, (0.4--0.85\,$\mu m$) and an infrared spectroscopic mapper, L'Ralph, (1.0--3.6\,$\mu m$) to analyze the composition of surface materials such as rock, ice, and organics. 
A thermal infrared spectrometer, L'TES (6--75\,$\mu m$), which will be used to investigate surface thermal properties and infer the rock-to-ice ratio and porosity of the outermost layers.
Observations from Lucy will therefore provide valuable tests of whether the structures predicted in our models, such as hydrated interiors or possible dehydration in central regions, are present in D-type asteroids.

Future laboratory and observational studies could provide concrete tests that may confirm or rule out the hypothesis that certain iron meteorites originated from icy planetesimals. 
For example, high-pressure and high-temperature experiments using mixtures of silicate, metal, and ice could determine whether metallic segregation occurs under water-rich conditions and whether Fe-Ni-S melts can migrate through hydrated silicate matrices. 
Measurements of Fe, O, and S isotopic compositions in iron meteorites could reveal signatures of oxidation or isotopic fractionation produced by interaction with water during metal-silicate differentiation. 
Furthermore, comparison of the inferred redox state (fO$_{2}$) and light-element contents (S, C, H) in metallic cores with model predictions could indicate whether differentiation occurred in hydrous or anhydrous environments. 
Upcoming spacecraft observations, such as those by the Psyche mission, may also test whether metallic bodies show magnetic or mineralogical evidence consistent with differentiation of volatile-rich planetesimals.

%
\section{Conclusions}
In this study, we developed a numerical model that comprehensively incorporates the complex processes occurring in icy planetesimals to provide a unified explanation of their diverse thermal evolution. 
We performed simulations varying the final radius, the onset timing and duration of growth, and the growth mode. 
Our results showed that larger final radii and earlier onset of growth since CAI formation lead to higher internal temperatures, which is consistent with previous studies.
The addition of growth and impact heating to the model make thermal evolution more complex.
For growth mode with $\beta=2$ (runaway growth), internal temperatures increase when growth is completed earlier. 
Regardless of when accretion ends, the interior tends to become isothermal, with only modest heating near the surface.
For $\beta=0$ (linear growth), internal heating and thermal structures vary significantly depending on the onset timing and duration of growth. When the growth begins earlier, internal heating is greater.
If the duration of growth is longer, the central region becomes hotter, whereas much of the planetesimal remains at temperatures below the melting point of ice.
Additionally, for a larger final radius, influence of the impact heating becomes stronger. 

These findings support the possibility that icy planetesimals may possess diverse internal structures, as suggested by observations of bodies believed to have originated from them.
In the case of linear growth, the radius increases from an early stage, which facilitates a temperature rise. 
Our results show that the constituent materials of Ryugu, which kept below 40${}^\circ$C, should have existed locally near the surface of the hydrated mineral layer. 
This region exists to a small extent even if the radius of the planetesimal that is Ryugu's parent body reaches several hundred kilometers. 
This scenario allows for a larger parent body than previously assumed (typically around 10 to several tens of kilometers; \citep[e.g.][]{travis2005,neumann2015,neumann2021}, because the high heat transport efficiency provided by convection within hydrated minerals and solid ice suppress the overall temperature increase.
In the runaway growth mode, if growth begins 2.0~Myrs after the CAI formation and is completed within 0.4~Myrs, a relatively broad region in the parent body with a radius of several hundred kilometers could be the Ryugu's constituent materials in terms of thermal conditions. 
However, maintaining high porosity under the pressures of such large bodies is problematic, and previous studies suggest that micro-porosity would be largely eliminated in the deep interior \citep[e.g.][]{neumann2021}. 
Therefore, preservation of porous materials may have been limited to shallower layers, which would then provide the main source of Ryugu's constituent materials.
Meanwhile, if the growth duration is prolonged, even aqueous alteration may not occur throughout the body.
Iron meteorites with isotopic compositions similar to those of carbonaceous chondrites may have undergone metal melting in the deep region if icy planetesimals that began growing 1.0\,Myrs after the formation of the CAI reach a radius of 200\,km or more within 0.4\,Myrs through linear growth. 
In other words, the possibility that the parent body of the iron meteorite is an icy planetesimal cannot be ruled out.
Furthermore, the evolution of the structure, which grows to a final radius of 250\,km and develops a hydrous core with a radius of 170--200\,km covered by a liquid water layer several tens of kilometers thick, may correspond to Saturn's moon Enceladus.
There are many additional processes in icy planetesimals that are not addressed in this study. 
Incorporating these processes will be necessary to further deepen our understanding of the thermal evolution of icy planetesimals.
\section*{Declarations}

\section*{Ethics approval and consent to participate}
Not applicable.

\section*{Availability of data and materials}
The datasets used and/or analyzed during the current study are available from the corresponding author upon reasonable request.

\section*{Competing interests}
There is no competing interest.

\section*{Funding}
This work was supported by JSPS KAKENHI grant numbers JP22K03700, JP17K05635 and JP22740285.

\section*{Authors' contributions}
Jun Kimura conducted data analysis and interpretation.
Ryusei Satoh and Jun Kimura developed the numerical model and performed calculations and analyses.
Kentaro Terada and Sho Sasaki supported interpretation.
The manuscript was written by all authors.

\acknowledgments{We sincerely thank the two anonymous reviewers for their constructive comments and helpful suggestions, which greatly improved the quality of this manuscript.
This study was supported by KAKENHI from the Japan Society for Promotion of Science (Grant No. JP22K03700 and JP22740285).
}
\clearpage
\clearpage
\clearpage
\clearpage
\clearpage
\clearpage
\clearpage
\clearpage
\clearpage
\clearpage
\clearpage
\clearpage
%
\clearpage
\bibliographystyle{elsarticle-harv} 
\bibliography{ref_r3}
\end{document}